\documentclass[sigconf]{acmart}
\AtBeginDocument{%
  }

\setcopyright{acmlicensed}
\copyrightyear{2018}
\acmYear{2018}
\acmDOI{XXXXXXX.XXXXXXX}
\acmConference[Conference acronym 'XX]{Make sure to enter the correct
  conference title from your rights confirmation email}{June 03--05,
  2018}{Woodstock, NY}
\acmISBN{978-1-4503-XXXX-X/2018/06}

\usepackage{graphicx}
\usepackage{placeins}
\usepackage{float}

\usepackage{booktabs}
\usepackage{tabularx}
\usepackage{array}

\begin{document}

\title{If My Toy Could Talk: How Young Children Imagine, Design, and Test AI-Enabled Toys}

\author{Feiwen Xiao}
\email{feiwenx@upenn.edu}
\affiliation{%
  \institution{University of Pennsylvania}
  \city{Philadelphia}
  \state{Pennsylvania}
  \country{USA}
}

\author{Ruiyang Wu}
\affiliation{%
  \institution{Graduate School of Education, University of Pennsylvania}
  \city{Philadelphia}
  \state{Pennsylvania}
  \country{USA}
}
\email{ruiyangw@upenn.edu}

\author{Xinyue Cui}
\affiliation{%
  \institution{University of Florida}
  \city{Gainesville}
  \state{Florida}
  \country{USA}
}

\author{Yasitha Rajapaksha}
\affiliation{%
  \institution{Department of Computer Science, North Carolina State University}
  \city{Raleigh}
  \state{North Carolina}
  \country{USA}
}
\email{yrajapa@ncsu.edu}

\author{Xiaoyi Tian}
\affiliation{%
  \institution{Department of Information Technology, Kennesaw State University}
  \city{Marietta}
  \state{Georgia}
  \country{USA}
}
\email{xtian5@kennesaw.edu}

\author{Shiyan Jiang}
\affiliation{%
  \institution{University of Pennsylvania}
  \city{Philadelphia}
  \state{Pennsylvania}
  \country{USA}
}

\author{Tiffany Barnes}
\affiliation{%
  \institution{Computer Science, North Carolina State University}
  \city{Raleigh}
  \state{North Carolina}
  \country{USA}
}
\email{tmbarnes@ncsu.edu}

\renewcommand{\shortauthors}{Trovato et al.}

\begin{abstract}
To investigate the design space where children might design AI chatbots for their own toys, we developed ToyTalk, a technology probe that positions children as designers of LLM-enabled toys. Children begin with a familiar toy, configure its AI-enabled version through a no-code interface, and then interact with and test the character. We deployed ToyTalk with 76 children aged 7--9 across five elementary schools in a two-week, teacher-led workshop. Using toy profiles, versioned configurations, and conversation logs, we examined children's character designs, conversational probes, and responses to observable mismatches. Within the workshop, children predominantly designed toys with socially positive personalities, supportive roles, and interpersonal rules. In conversation, they most often probed identity and knowledge, while also testing capabilities, memory, and relationships. When mismatches arose, children typically responded through correction, persistence, and retesting, while few returned to reconfigure the system. We discuss implications for children’s design agency, testing practices, and expectations of coherence in child-facing generative AI.
\end{abstract}

\ccsdesc[500]{Human-centered computing~Empirical studies in HCI}
\ccsdesc[500]{Human-centered computing~Interactive systems and tools}
\ccsdesc[300]{Applied computing~Interactive learning environments}
\ccsdesc[300]{Information systems~Information systems applications}

\keywords{Children, Chatbots, Children as designers, Generative AI, elementary school, K-12 education}

\begin{teaserfigure}
    \centering
    \includegraphics[width=0.90\textwidth]{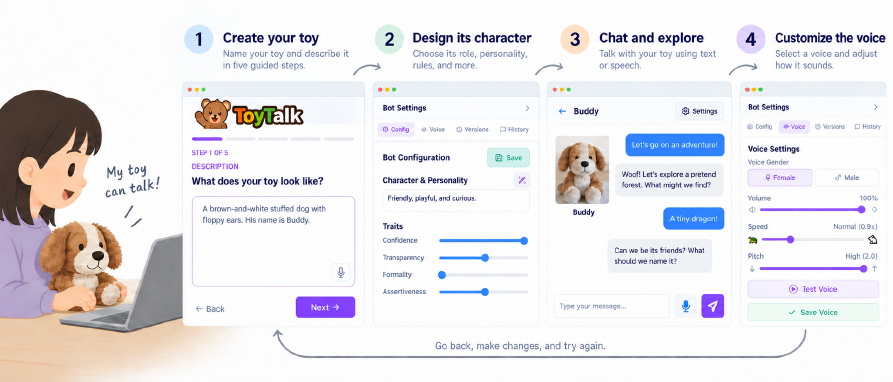}
    \caption{ToyTalk positions children as designers of AI-enabled versions of their familiar toys. Through a no-code interface, children describe a toy, shape its character and behavioral traits, converse with it through text or speech, and customize its voice. Children can return to revise their settings. Interface illustrations are adapted from the implemented system; the scene, configuration, and dialogue are illustrative.}
    \Description{Illustration of a child and a stuffed dog beside four connected ToyTalk interface panels. At left, the child sits with the dog at a laptop; the same dog appears on screen as Buddy. From left to right, the panels show the first of five guided setup questions; a character described as friendly, playful, and curious with four trait sliders; a conversation in which Buddy proposes a pretend forest adventure and the child imagines a tiny dragon; and voice selection with volume, speed, pitch, test, and save controls. Arrows connect the panels, and a long return arrow beneath them points back to setup.}
\label{fig:toytalk-teaser}
\end{teaserfigure}

\received{20 February 2007}
\received[revised]{12 March 2009}
\received[accepted]{5 June 2009}

\maketitle

\section{Introduction}
A stuffed animal can be a friend, a doll can be someone to care for, and a toy car can become a partner in an imagined adventure. Through play, children routinely give toys roles, personalities, and meanings beyond their literal functions \cite{palaiologou2021make}. When computation is added to a toy, it can reshape what the object affords in play\cite{mertala2020connectivity}. Recent advances in large language models (LLMs) introduce a new possibility: these imagined characters can now talk back. LLM-enabled toys can participate in open-ended conversation, generate stories and other content, and respond dynamically to what children say. Such capabilities are already appearing in commercial toys and companions enabled by artificial intelligence (AI)\cite{xiao2025intelligent,dangol2026toys}. As generative AI moves from screens into children's play objects, however, an important question emerges: \textit{who gets to decide what these AI characters should be?}

Although most emerging AI-enabled toys have predefined identities, personalities and ways of interacting, recent work on LLM-enabled toys showed that children do not simply accept these characters as given. They actively ask who the toy is, test what it can do, probe its boundaries, and challenge behavior that does not meet their expectations\cite{dangol2026toys}. Earlier work with conversational agents embodied in children's stuffed animals likewise suggests that children care about whether an agent's voice and behavior fit the character they imagine for the physical toy, and highlights the importance of giving children greater agency over these representations \cite{hubbard2021voice}. These observations motivate a different design possibility: rather than interacting with an AI-enabled toy designed for them, what if children could decide what their own toy becomes with AI?

Growing evidence shows that children can meaningfully participate in shaping AI building, training, and experimenting with intelligent systems \cite{williams2019popbots,druga2021perceptions}, and creating their own conversational agents \cite{tian2023amby}. With LLMs, children can make high-level decisions about an agent's character and behavior, specifying chatbot characters, topics, rules, constraints, and knowledge sources and then iteratively testing the resulting agent \cite{tian2026hallucination}. Creating their own artifacts may also increase children’s valuation of them: as “IKEA effect” indicated, young children value self-created objects more than equivalent objects created by others \cite{norton2012ikea}. Designing and testing AI systems may also deepen children’s understandings of how they work and fail\cite{williams2019popbots,druga2021perceptions,tian2026hallucination}. However, designing an LLM-enabled toy introduces a distinct challenge: children's design choices do not fully determine how the resulting character will behave. The model may elaborate aspects of the character that were never specified, interpret desired properties in unexpected ways, or fail to sustain properties that children expect to remain stable. Thus, once an AI toy becomes generative, children must not only decide what they want the toy to be, but also encounter and make sense of how the model realizes those design choices in interaction.

To investigate this design space, we developed ToyTalk, a technology probe that positions children as designers of LLM-enabled toys. Children begin with a familiar physical toy, configure what its AI-enabled version should be like, and then interact with and test the resulting character. Through a no-code interface, they specify the toy’s envisioned role and capabilities, personality, and behavioral boundaries, while retaining the ability to revise these choices. We deployed ToyTalk with 76 second- and third-grade children across five elementary-school sites. Rather than treating design and subsequent interaction as separate stages, we examine how children move between defining what their toy should be, probing what it has become, and responding when its conversational behavior diverges from their expectations. We ask:
\begin{itemize}
\item {\texttt{RQ1}}: What familiar toys did children bring into the classroom, and how do they envision what their familiar toys could be like when brought to life with AI?
\item{\texttt{RQ2}}: Once their toy is configured, what and how do children probe and test its behavior?
\item{\texttt{RQ3}}: When tensions arise between children's expectations for their toys and its actual behavior, how do children respond? 
\end{itemize}

Through this work, we contribute ToyTalk, a technology probe that gives young children accessible control over the character and behavior of AI-enabled versions of familiar toys. Empirically, we characterize how children design these characters, probe their emerging behavior, and respond when that behavior diverges from their expectations. Building on these findings, we identify implications for supporting children’s agency as designers of generative AI, their emerging practices for testing AI behavior, and the design of AI characters that remain coherent across interaction.

\section{Related Work}
\subsection{Children's interaction with conversational AI}
Children nowadays are interacting with conversational AI on a daily basis. They talk to voice assistants (e.g., Amazon Alexa and Google Home), interact with intelligent agents embedded on learning platforms (e.g., Khan Academy), or even play around with generative AI systems (e.g., ChatGPT). Prior work shows that children actively make sense of these systems, attributing intelligence, personality, and social qualities to conversational agents \cite{druga2017hey,xu2020you}. Children also use conversational agents as information sources, although how they formulate questions and respond to answers varies with age and prior experience \cite{lovato2019hey,orancc2021alexa}. When interactions break down, children repeat, reformulate, or otherwise adapt their speech, gradually learning what an agent can understand and do \cite{cheng2018doesn,beneteau2019communication}.

Generative AI broadens these interactions by enabling more open-ended forms of engagement. Studies of children and adolescents using generative chatbots have identified testing, socializing, and content-exploring behaviors, including deliberately asking questions with known answers to probe what the AI knows \cite{belghith2024testing,vahedian2025ask}. In longer-term use, children also evaluate and adapt to AI responses: Oh et al. found that children engaged in wide-ranging, curiosity-driven information seeking and adjusted their follow-up behavior based on the perceived accuracy of the AI's responses \cite{oh2025hey}. At the same time, children's understandings of generative AI can develop rapidly and vary with how the system is presented. Kosoy et al. found differences in children's interactions across text and image generation modalities \cite{kosoy2024children}, while Xu et al. showed that speech and embodiment shaped children's perceptions of a generative AI partner \cite{xu2025makes}.

This line of work positions children as active interpreters of conversational AI who probe its capabilities, evaluate its responses, and adapt through interaction. Yet these agents are typically designed in advance, with their identity, role, and interactional framing largely predetermined. We know less about how children interact with conversational AI when they can help define who the agent is and encounter it through a familiar play object, motivating closer attention to customizable AI-powered toys.

\subsection{Children's Play and AI-Enabled Toys}

AI has long been incorporated into toys to make them more responsive and interactive, extending the possibilities of children's play. Early robotic toys such as AIBO elicited reciprocal and social responses from children \cite{kahn2004robotic, melson2009robotic}, while conversational toys such as Hello Barbie\footnote{\url{https://m.service.mattel.com/us/Technical/productDetail?prodno=DKF74&siteid=27}} and CogniToys Dino\footnote{\url{https://www.kickstarter.com/projects/cognitoys/cognitoys-internet-connected-smart-toys-that-learn}} enabled children to speak directly with toy characters. Prior work showed that children formed expectations about these toys' intelligence and conversational abilities \cite{mcreynolds2017toys}, and that adding connectivity and speech could reshape the affordances of otherwise familiar toys \cite{mertala2020connectivity}. Researchers have also explored AI-enabled toys for learning. For example, Maya used image processing and natural language processing to recognize children and adapt concept-learning activities to their learning pace \cite{akdeniz2021maya}. These earlier systems varied in the conversational interactions they supported. Hello Barbie, for example, used speech recognition to select from more than 8,000 prerecorded lines of dialogue rather than generating open-ended responses \cite{steeves2020dialogic}.

Recent generative AI substantially expands this design space by enabling toys to sustain open-ended dialogue, generate stories and other content, and adapt their responses to conversational context. These capabilities are increasingly appearing in commercial products such as Curio's Grok, Grem, and Gabbo,\footnote{\url{https://heycurio.com/}} FoloToy's conversational plush toys,\footnote{\url{https://folotoy.com/}} and Haivivi's BubblePal.\footnote{\url{https://www.haivivi.com/}} Yet empirical research on children's actual interactions with generative AI toys remains limited. One exception is a recent study from Dangol et al. \cite{dangol2026toys}. They found that children actively probed commercial GenAI toys to understand their identities, capabilities, and boundaries, with mismatches sometimes prompting further testing or adversarial play . Related work with younger children has similarly documented challenges around conversational turn-taking, emotional responsiveness, and integrating GenAI toys into pretend play \cite{goodacre2026ai}. Overall, these studies begin to show how children make sense of generative AI toys, but this emerging context remains underexplored.

Moreover, children in prior studies have largely interacted with toy characters whose identities and personalities were designed in advance. Some work has begun to explore conversational agents embodied in children's existing stuffed animals \cite{hubbard2021voice}, suggesting that familiar toys carry pre-existing meanings and expectations for children. However, children have generally had limited opportunities to define the conversational character itself. 

\subsection{Children as Designers of AI Systems}

Beyond interacting with AI, prior work has increasingly positioned children as designers who can build and shape intelligent systems themselves. Constructionist environments such as PopBots \cite{williams2019popbots} and Cognimates \cite{druga2021perceptions} have enabled children to program, train, and experiment with AI behaviors, while Zhorai \cite{lin2020zhorai} allowed children to train a conversational agent through interaction. Other work has given youth more direct control over the design of conversational AI. AMBY \cite{tian2023amby}, for example, supports youth in creating personally meaningful conversational agents through training data and conversational flows. More recent LLM-based systems further lower the barrier between children’s ideas and functioning applications. For instance, Aptly \cite{patton2025aptly} translates natural-language descriptions into executable applications, while Tian et al. \cite{tian2026hallucination} enabled middle-school students to configure and iteratively test LLM-powered chatbots by specifying their characters, topics, rules, constraints, and knowledge sources. The literature demonstrates that children can make consequential decisions about what AI systems should do and how they should behave.

A parallel body of child-centered and participatory design research has involved children in envisioning what AI and embodied agents should be like. Children have contributed design concepts for future AI systems \cite{bai2023participatory,wilson2025childcentred}, co-designed social robots and their interaction scenarios \cite{alvesoliveira2021children,neerincx2023codesigning}, and expressed preferences concerning robots' personalities, roles, and social behavior \cite{levinson2025think}. Zhou et al., for example, engaged young children in co-designing characters and social-emotional scenarios to elicit desired appearances, roles, and behaviors for storytelling agents \cite{zhou2024lollipop}. Related work on the Internet of Toys has likewise demonstrated the value of involving children directly in the design of connected toys \cite{vanmechelen2019internet,yamadarice2019including}. These studies establish children not simply as recipients of child-facing technologies, but as participants capable of articulating how intelligent and embodied systems should act in their lives. Generative AI has further expanded the forms of design and creative agency available to children. A recent review found child--AI co-creation research concentrated particularly in storytelling and visual creation, with AI acting as a generator, facilitator, collaborator, or enhancer \cite{cai2025childai}. Systems such as Mathemyths \cite{zhang2024mathemyths}, ChatScratch \cite{chen2024chatscratch}, and other generative creativity tools allow children to shape stories, programs, images, and other creative artifacts with AI \cite{zhang2024mathemyths,chen2024chatscratch, newman2024darth}. This work provides important evidence about children's control and agency in generative processes, but the object being designed is typically the resulting creative artifact rather than the AI system itself.

In general, the literature establishes that children can design AI systems, contribute to the design of embodied agents and toys, and exercise creative agency with generative AI. However, we know comparatively less about children designing \textit{LLM-enabled AI toys} that they subsequently interact with themselves. This setting brings these strands together: children must decide what a familiar toy should be able to do, what kind of character it should become, and how it should behave, while the generative system can subsequently elaborate on or depart from those design choices through interaction. Our work examines this intersection by giving children design agency over what their own familiar toys could become with generative AI and studying how those designs are subsequently interpreted, tested, and negotiated in interaction.

\section{Method}
\subsection{Designing ToyTalk as a Technology Probe}
We developed ToyTalk as a \textbf{technology probe} to explore how children might design and interact with LLM-enabled versions of familiar toys (Figure~\ref{fig:toytalk_interface}). Technology probes are simple and flexible technologies introduced early in a design space to understand users and practices in context, explore emerging technological possibilities and limitations, and inspire future design \cite{hutchinson2003technology}. Rather than approximating a finished product, ToyTalk made the possibility of designing an LLM-enabled toy concrete: children could define what their toy should be like, experience how those choices were realized through generative conversation, and revisit their design in response. This allowed us to examine both what children imagined for their toys and how they interpreted and responded to the behavior that emerged from their designs. 

\subsubsection{Design Goals}
Three \textbf{design goals (DG)} guided the design of ToyTalk:

\begin{enumerate}
    \item \textbf{DG1. Anchor AI design in children's existing play worlds.} We wanted children to approach AI design through something already concrete and meaningful to them rather than an abstract chatbot or predefined AI character. ToyTalk therefore centers the activity on a familiar physical toy, providing a starting point for imagining what AI might add while leaving open whether children preserve, extend, or transform what the toy can be.
    \item \textbf{DG2. Lower the barrier to meaningful AI design.} We sought to give elementary-aged children consequential control over an LLM-enabled character without requiring programming or prompt-engineering expertise. ToyTalk therefore translates AI configuration into decisions children can make about the toy itself---what it should do, what it should be like, and how it should behave---while ensuring that these choices shape the behavior they subsequently encounter.
    \item \textbf{DG3. Support open-ended exploration and iterative refinement. }Because configuration cannot fully determine the behavior of an LLM-enabled character, we wanted children to experience how their design choices were realized in interaction and continue shaping the toy in response. ToyTalk therefore supports open-ended conversation while keeping children's designs editable, allowing them to move between configuration, interaction, revision, and retesting as they choose.
\end{enumerate}

As a technology probe deployed within a classroom curriculum, ToyTalk also included instructor-facing affordances to support specific learning activities, which we describe below.

\paragraph{Bringing a Familiar Toy into ToyTalk.}
Children begin by uploading a picture of the physical toy they brought to the activity, giving it a name and optionally a short description. The toy's image remains visible during subsequent interaction, visually linking the conversational character to the physical object from which it originated. Rather than beginning with a predefined conversational character, ToyTalk centers the design process on a familiar object that may already carry meanings and expectations for the child.

\paragraph{Providing Children with Levers to Shape the Toy.}
After creating the toy, children use a guided, no-code configuration process that translates LLM settings into child-facing questions, such as “What is your toy’s job or special purpose?” Through open-ended responses, children specify the toy’s role and capabilities, personality, and behavioral boundaries, which are incorporated into the character specification guiding subsequent responses. They can also add knowledge sources that the toy can draw on during conversation and customize its voice type, volume, speaking speed, and pitch. Together, these features provide accessible levers for shaping both the toy’s behavior and the information it can use.

\paragraph{Experiencing the Designed Toy Through Conversation.}
Once configuration is complete, children can enter an open-ended conversation with their LLM-enabled toy through text or speech, with generated responses spoken aloud using the configured voice. For each turn, ToyTalk generates a response based on the toy's current character specification and conversational context. The interface does not prescribe what children should ask: they can ask questions, request stories or performances, provide information about themselves or the toy, or pursue topics of their own choosing. This open-ended interaction allowed children's own ways of exploring and making sense of the character to emerge.

\subsubsection{Probe Design} 
\begin{figure}[H]
    \centering
    \includegraphics[width=\linewidth]{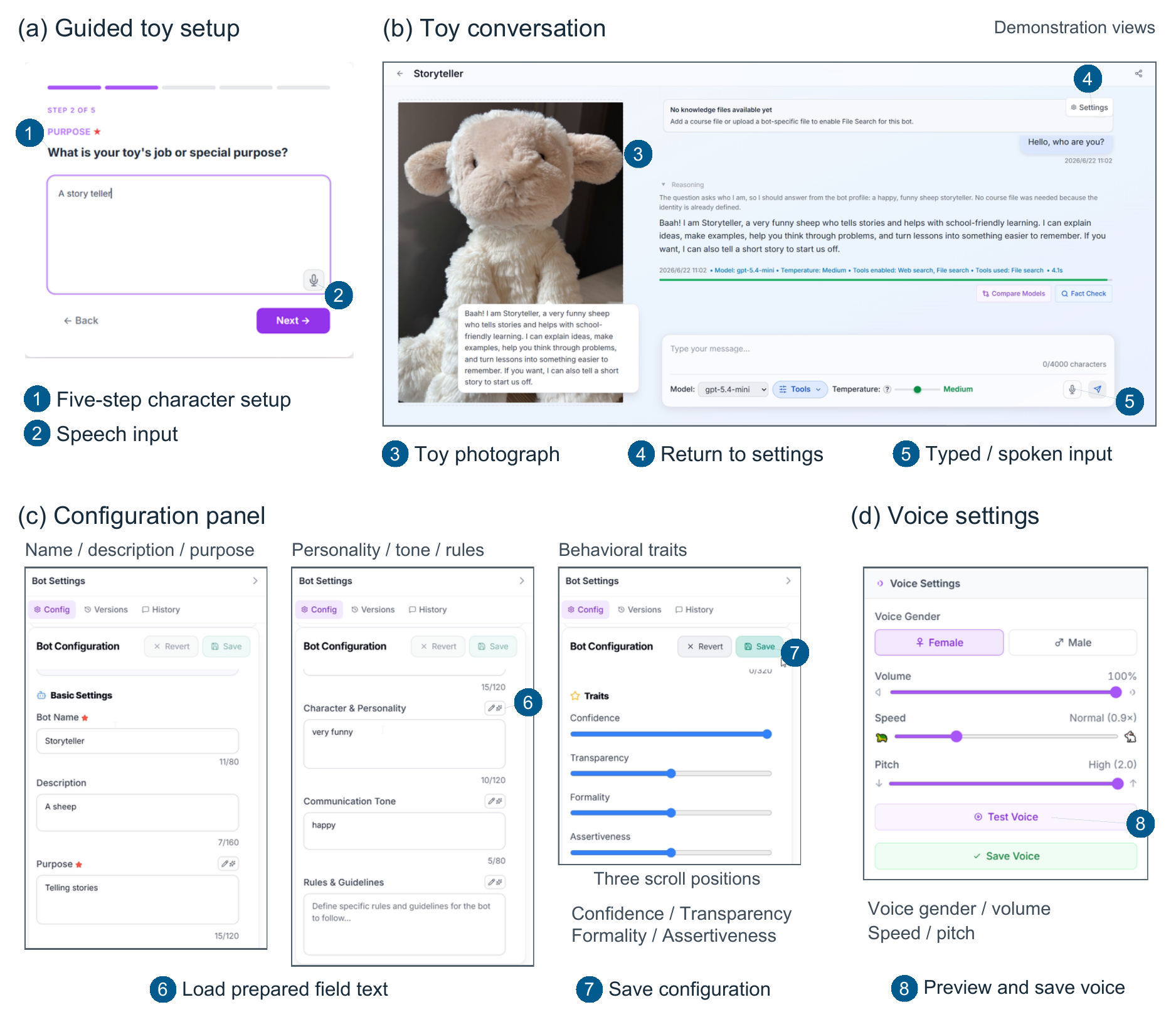}
    \caption{ToyTalk interface views from a demonstration. (a) Guided setup presents a purpose question and supports speech input. (b) The uploaded toy photograph appears beside the conversation; Settings provides access to revision. (c) Three separate scroll positions show editable character fields and the Confidence, Transparency, Formality, and Assertiveness controls. Prepared-text buttons fill individual fields, and Save commits changes. (d) Voice settings control voice choice, volume, speaking speed, and pitch, with preview and saving. Screens show demonstration content rather than participant interaction records; the displayed model and knowledge-base state belong to the demonstration.}
    \label{fig:toytalk_interface}
    \Description{Four demonstration panels arrange ToyTalk's setup and chat views above its configuration and voice controls. Panel a has a five-step progress indicator, a purpose text box, and a microphone. Panel b places a plush sheep photograph beside the conversation, with Settings at the upper right and message input below. Panel c shows three adjacent scroll positions for identity and purpose, personality and rules, and four behavioral sliders. Panel d contains voice selection, volume, speed, pitch, preview, and save controls. Numbered markers identify: one, character setup; two, speech input; three, toy photograph; four, settings; five, typed or spoken chat input; six, prepared field text; seven, configuration saving; and eight, voice preview and saving.}
\end{figure}

\paragraph{Keeping the Toy Open to Revision.}
Children can reopen the toy's settings at any point to modify its configuration or voice and then return to conversation. Saved configurations are retained as versions, preserving earlier states while allowing changes to persist. ToyTalk does not prescribe when children should revise, allowing them to move between conversation and reconfiguration as they choose. This flexibility allowed us to observe when interactions with the generated character prompted children to revisit earlier design decisions. Figure~\ref{fig:toytalk_interface} presents the main interface views.

\paragraph{Supporting Instructor-Orchestrated Encounters with AI Errors.}
The workshop is situated in the classroom; ToyTalk included an instructor-controlled mechanism beyond its child-facing design features, to introduce factual errors in selected responses. This probe-specific affordance allowed instructors to create opportunities for children to encounter and scrutinize unreliable AI-generated information rather than relying on such errors to arise naturally.

\begin{figure}
    \centering
\includegraphics[width=\linewidth]{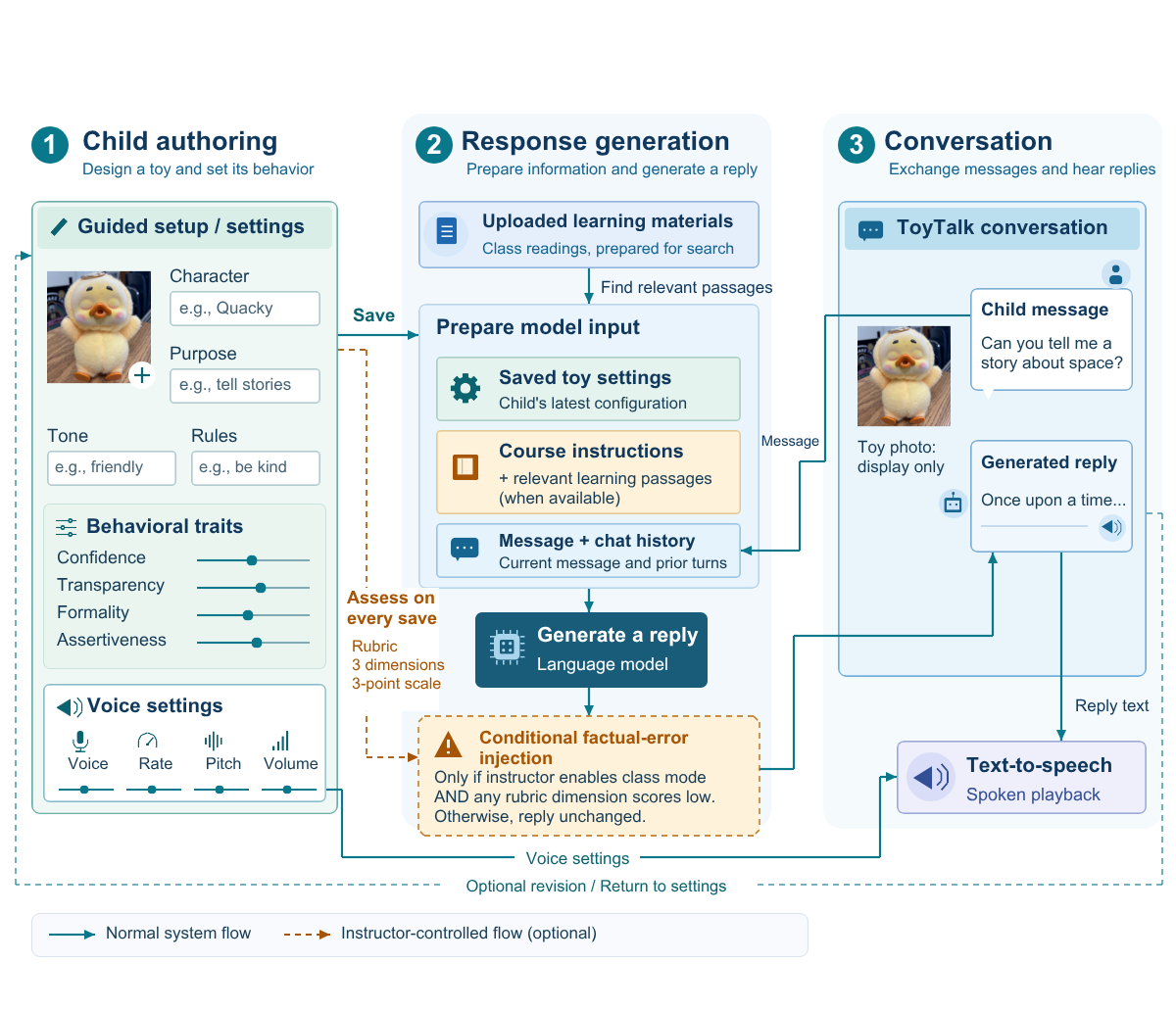}
  \caption{ToyTalk system workflow. Children configure their toy's character instructions and behavioral traits. ToyTalk generates replies from the saved configuration, course instructions, relevant learning passages when available, and the current message and chat history. Voice settings control spoken playback, while the toy photograph is for display only. Orange dashed elements indicate the optional instructor-controlled factual-error injection pathway: injection occurs only when classroom mode is enabled and at least one rubric dimension scores low. The teal dashed path indicates an optional return to settings. Example settings and dialogue are illustrative.}
    \Description{Three columns connect child authoring, response generation, and conversation. Solid arrows bring saved toy settings, course instructions with relevant retrieved passages, and the current message with chat history together as inputs to a language model. Each configuration save also triggers assessment using a rubric with three dimensions and three scoring levels, shown along an orange dashed route. After generation, a conditional step introduces factual errors only if the instructor has enabled class mode and at least one dimension scores low; otherwise, the reply passes unchanged. The reply appears beside a toy photograph marked for display only. Separate arrows carry reply text and voice settings to spoken playback. A teal dashed loop returns from conversation to settings for optional revision.}
\label{fig:toytalk_architecture}
\end{figure}

\subsubsection{System Implementation}
ToyTalk uses a service-based architecture separating user management, toy configuration, knowledge processing, and conversation generation. A React frontend communicates with Python/FastAPI services through nginx. The user service manages accounts and classrooms, the bot service stores versioned toy configurations, and the chat service orchestrates retrieval and LLM inference.

Children's saved configuration, including purpose, character, tone, rules, and behavioral traits, is incorporated into the system prompt. For each message, ToyTalk combines this configuration with classroom context, recent conversation history, and relevant retrieved knowledge. Uploaded materials are chunked and embedded using \nolinkurl{sentence-transformers/all-MiniLM-L6-v2}, with relevant passages retrieved from PostgreSQL/pgvector using cosine similarity.

Responses are generated by a self-hosted \nolinkurl{meta-llama/Llama-3.2-3B-Instruct} model served through vLLM. The response-generation agent can invoke knowledge-base search, a restricted calculator, or optional web search. Responses are stored and streamed to the browser. The toy image is displayed locally but not provided to the model. Speech input and output use the browser Web Speech API, with playback controlled by the child's selected voice, rate, pitch, and volume. Additional implementation details are provided in Appendix~\ref{sec:system-implementation-details}.

\paragraph{Instructor-Controlled Hallucination Injection.}
ToyTalk included a class-level mode through which instructors could deliberately introduce factual errors into selected toy responses. In our study, this mode was enabled on Week 2, Day 3, immediately before the AI-mistakes lesson, which asked children to verify AI-generated information against the reading material. Each time a child saved a configuration, a locally hosted LLaMA-3B model evaluated it using a three-dimension, three-point rubric (Appendix~\ref{app:configuration_rubric}). Configurations receiving a low score on any dimension were eligible for error injection. For eligible configurations, an LLM altered selected factual details, such as numbers, times, locations, or names, before the response was shown to the child; for example, ``The caterpillar ate \textit{three} apples'' could become ``The caterpillar ate \textit{five} apples.'' Injection ceased once a revised configuration received no low rubric scores. These errors were intentionally introduced for instructional purposes and should not be interpreted as spontaneous failures of the underlying response-generation model.

The study took place during the first two weeks of a three-week, in-person summer-school program at five Title I elementary schools in a public school district in the southeastern United States. The program integrated English Language Arts (ELA) with introductory AI-literacy activities. Seventy-eight second- and third-grade students aged 7--9 participated, including 45 second-graders and 33 third-graders. One family declined research participation, and one student attended only the first week and did not complete the toy-design activities, yielding an analytic sample of 76 children. Table~\ref{tab:study_sites} reports school-level racial and ethnic composition. Teachers described multilingual learners as a substantial part of some classrooms and noted students' familiarity with school computers and typing. Students accessed ToyTalk using school-provided computers. The study was approved by the relevant institutional review board and followed approved consent procedures.

Three elementary educators led instruction across the five sites: one taught at one site, while the other two each taught at two. Teachers retained responsibility for instruction, pacing, and classroom management, while the research team provided pre-program professional development, daily debriefs, and implementation planning. Teacher feedback informed adjustments to worksheets, pacing, modeling, and other instructional supports.

Session structure varied across sites: one used approximately 45-minute grade-specific sessions, two used approximately 60-minute combined Grade 2--3 sessions, and two used approximately 30-minute grade-specific sessions. Teachers adapted modeling, written work, peer support, and activity time to local classroom contexts.

\subsection{Procedure} 
\label{sec:procedure}

\begin{figure}[h]
    \centering
    \includegraphics[width=\linewidth]{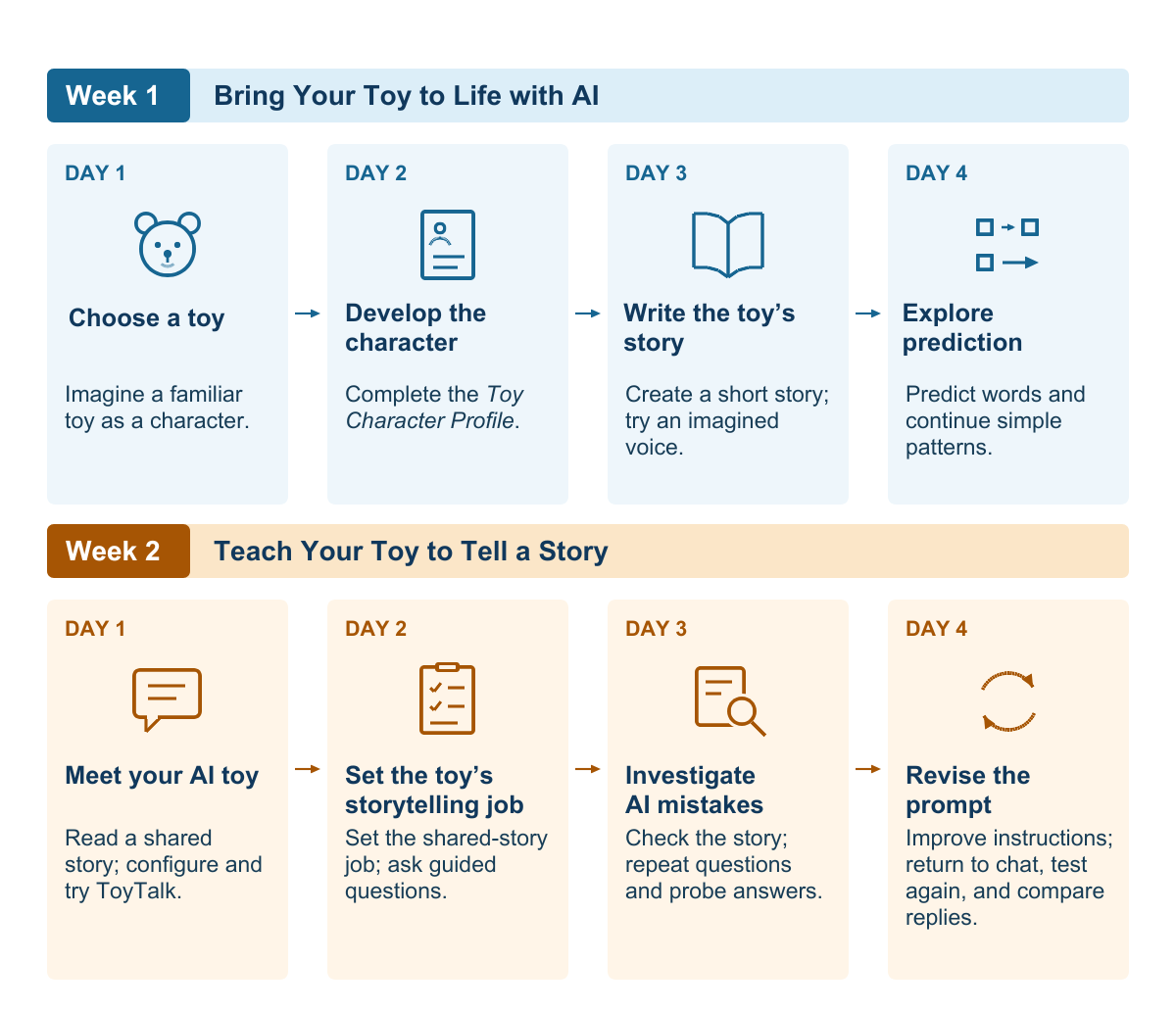}
    \caption{Intended activity sequence during the first two weeks of the ToyTalk summer-school program. Children developed familiar toys into AI-enabled characters and probed their behavior throughout the activities. They first explored next-word prediction, then configured and iteratively tested their toys in ToyTalk. In Week 2, they used \textit{The Very Hungry Caterpillar} as a shared reference to probe responses, investigate AI mistakes, and revise their configurations. Teachers led the activities with research-team support; pacing and completion varied across schools.}
\label{fig:toytalk-procedure}
    \Description{Two rows present four days each, with Week 1 above Week 2 and arrows running left to right. Week 1 progresses through choosing a familiar toy, completing its character profile, writing a short story and trying an imagined voice, and exploring word prediction and patterns. Week 2 begins with reading a shared story, configuring the toy, and trying ToyTalk. Day 2 emphasizes teacher-guided testing through story-related and off-topic questions. On Day 3, children check the story, repeat questions, and probe answers to investigate artificial intelligence mistakes. Day 4 involves improving instructions, returning to chat, testing again, and comparing replies. Each day has an activity icon, task heading, and short explanation.}
\end{figure}

The two-week workshop followed two broad phases. In Week 1, children developed the character of a familiar toy and were introduced to basic ideas about how generative AI produces language. In Week 2, children brought their toy designs into ToyTalk and used a shared storytelling activity to explore, evaluate, and revise the toy's generated behavior. Figure~\ref{fig:toytalk-procedure} summarizes the intended activity sequence. To support implementation, the research team met with the three teachers daily to review upcoming activities, provide instructional materials, and offer guidance on using ToyTalk.

\paragraph{Week 1: Bring Your Toy to Life with AI.}
Children chose a familiar toy and developed it as a story character using a \textit{Toy Character Profile}. Supported by word banks, they described its special job, personality, communication style, and things it could or could not do. They then wrote a short story featuring the toy and imagined or acted out how it would speak. Activities on next-word prediction and simple patterns introduced the idea that generative AI produces language by predicting likely continuations. Finally, teacher invited students to configure and test their own toy profile in ToyTalk.

\paragraph{Week 2: Teach Your Toy to Tell a Story.}
Children read \textit{The Very Hungry Caterpillar} and used their profiles as references to configure and try their toys in ToyTalk. They gave the toys a storytelling job centered on the book and asked guided test questions. Teacher uploaded the story to the knowledge base, which provided a reference for checking generated responses. An ``AI Mistake Detectives'' activity encouraged children to consult the story, repeat questions, notice how confident the AI sounded, and ask for explanations. A ``better prompt'' activity introduced \textit{role}, \textit{job}, \textit{tone}, and \textit{rules}, inviting children to revise their instructions and test again. Throughout the week, children could also ask their own questions to test the toys and return to settings as they wished.

\subsection{Data Collection}
We collected children’s toy-profile worksheets and system-generated ToyTalk logs. Worksheets captured children’s initial toy designs and served as a triangulation source for their digital profiles. ToyTalk automatically logged children’s conversations and configuration changes. The conversation logs contained 5,405 child and AI messages, with each record including message content, conversational order, timestamp, and associated toy configuration. Configuration logs preserved successive versions of each toy’s character or personality description, purpose, tone, behavioral rules, and conversational and voice settings, along with timestamps. These versioned records allowed us to trace changes to children’s toy designs over time and relate them to preceding and subsequent interactions.

\subsection{Participants and Settings}

\begin{table*}[!htbp]
\centering
\caption{Summer-program enrollment and school-level racial and ethnic composition across the five participating school sites. Enrollment counts describe program enrollment; demographic percentages describe the overall school populations.}
\Description{The table separates program enrollment from whole-school demographic percentages across five sites. Rows list Sites A through E and an enrollment total. One column group gives second-grade, third-grade, and total enrollment; another lists Hispanic, Black, White, two or more races, and Asian percentages. Enrollment totals 78 children: 45 in second grade and 33 in third grade. Demographic percentages describe the schools, not these children. Some demographic cells contain dashes; no combined demographic percentages are reported.}
\label{tab:study_sites}

\begin{tabular}{lrrrccccc}
\toprule
& \multicolumn{3}{c}{\textbf{Summer-Program Enrollment}} 
& \multicolumn{5}{c}{\textbf{School-Level Demographics (\%)}} \\
\cmidrule(lr){2-4}
\cmidrule(lr){5-9}

\textbf{Site} &
\textbf{Grade 2} &
\textbf{Grade 3} &
\textbf{Total} &
\textbf{Hispanic} &
\textbf{Black} &
\textbf{White} &
\textbf{Two or More Races} &
\textbf{Asian} \\
\midrule

Site A & 13 & 8 & 21 & 52.4 & 21.4 & 20.7 & 4.1 & -- \\
Site B & 9 & 8 & 17 & 6.3 & 46.8 & 31.1 & 11.1 & 4.7 \\
Site C & 10 & 4 & 14 & 67.1 & 14.9 & 12.2 & 4.1 & 1.1 \\
Site D & 3 & 5 & 8 & 20.6 & 16.9 & 52.8 & 6.9 & 2.8 \\
Site E & 10 & 8 & 18 & 37.1 & 3.7 & 51.5 & 7.0 & -- \\

\midrule
\textbf{Total} & \textbf{45} & \textbf{33} & \textbf{78} &
-- & -- & -- & -- & -- \\
\bottomrule

\end{tabular}
\end{table*}

\subsection{Data Analysis}
\subsubsection{RQ1. Toy Profile Analysis}
We used inductive qualitative content analysis supplemented by descriptive quantitative summaries to examine how children envisioned familiar toys brought to life with AI. Our analysis proceeded in three stages. First, we coded the toys children brought to the workshop based on images and profile information. Toy type could be identified for 71 of 76 cases (93.4\%), while the remaining five were retained in analyses of children's design descriptions when sufficient profile or configuration data were available.
Second, we open-coded children's descriptions of their toys, staying close to their language before iteratively grouping related codes into focused categories of \textit{roles and responsibilities}, \textit{personality characteristics}, and \textit{behavioral boundaries}. Toy Character Profiles were the primary source; when unavailable or unreadable, we used the child's earliest ToyTalk configuration to recover information. Third, we examined whether these patterns varied across toy types by cross-tabulating focused codes with the 71 identifiable toy types and calculating counts and within-type percentages. Quantitative summaries were used to characterize prevalence and distribution. 

\subsubsection{RQ2. What and how children tested their AI toys}
To examine what and how children tested their AI-enabled toys after configuration, we conducted a multiple-stage analysis of the conversation logs. We first identified testing probes, defined as child turns or utterances that elicited, checked, challenged, or otherwise made observable a property of the configured toy. We treated these probes as conversational epistemic actions that are used to obtain information about properties that are not directly observable \cite{kirsh1994distinguishing}. Greetings and related social talk, procedural talk, and other turns without an exploratory or evaluative function were excluded from this round of analysis. We also marked probes attributable to workshop scaffolding separately from children's self-generated probes when their provenance could be determined.

\paragraph{Understanding Testing Targets}
To characterize \textit{what} children tested, two authors independently reviewed 20\% of the conversation logs and open-coded the aspect of the toy examined in each probe, informed by prior work on child--agent interaction \cite{lovato2019hey,festerling2020alexa}. Initial codes were kept relatively fine-grained, distinguishing, for example, factual from mathematical knowledge, identity and role from preferences and traits, and generative from embodied or functional capabilities. Through affinity diagramming, the authors iteratively clustered conceptually related codes into broader \textit{testing targets} while retaining recurring distinctions as \textit{sub-targets}, corresponding to the ``aspects probed'' in Table~\ref{tab:test_targets}. Disagreements and ambiguous cases were discussed with reference to the surrounding conversational context until consensus was reached, with coding memos documenting category definitions, boundary cases, and representative examples. The authors then applied the refined codebook to the remaining logs while monitoring for additional categories; no new categories emerged. Any subsequent coding discrepancies were resolved through discussion.

\paragraph{Analyzing Testing Strategies} 
We next examined \textit{how} children made properties testable through conversation. Each probe was coded for its \textit{interactional form}, capturing the conversational action used to elicit evidence from the toy, such as asking an information or verification question, requesting a capability or performance, assigning a property, or presenting a hypothetical or challenge. We treated testing target and interactional form as analytically distinct because the same target could be probed through different actions, while the same form could address different properties. This distinction was informed by multidimensional dialogue analysis separating what an utterance addresses from the communicative action it performs \cite{bunt-etal-2012-iso}.

Finally, we examined children's probes as part of unfolding interaction rather than as isolated utterances. Drawing on interaction analysis and situated accounts of action,  which emphasize interpreting actions through their sequential organization and the local resources available within an unfolding activity \cite{jordan1995interaction,suchman1987plans,goodwin2000action}, we reconstructed each child's trajectory of configuration and testing, documenting configuration episodes, the type of change introduced, and the toy profile in effect when subsequent probes occurred. We considered together recurring \textit{Target $\times$ Form} co-occurrences, turn-by-turn child--AI exchanges, the active toy configuration, and each probe's position within the broader interaction trajectory.

We then traced how children continued or adapted tests after observing the toy's response, for example by repeating or varying probes, requesting demonstrations, correcting the toy, escalating requests, or supplying information and later testing its retention. Through iterative comparison of these multi-turn patterns within and across children, we identified higher-level \textit{testing strategies}: recurring ways children organized conversational actions to generate and evaluate evidence about their configured toys.




\subsubsection{RQ3. Analyzing Tensions and Children's Responses}
Two authors identified episodes in which the AI toy's behavior diverged from what children were probing, requesting, or attempting to establish. To avoid inferring expectations solely from researchers' judgments, we required observable evidence of the child's intended or expected behavior: (1) properties specified in the toy's configuration, (2) properties established in prior conversation, (3) direct requests or instructions, (4) expectations grounded in the toy's prior responses, or (5) subsequent child turns explicitly marking a response as incorrect, inconsistent, or undesired. Episodes began with the probe or action that made the expectation relevant and extended through children's attempts to address, retest, or abandon the tension. Cases without such evidence were excluded.

For each episode, we examined configuration logs for subsequent changes addressing the mismatch and coded these as system-level adaptations. We then inductively coded the \textit{tension type} and children's responses, iteratively developing a multi-label codebook with category definitions and representative examples. After finalizing the codebook, two coders independently coded the remaining episodes and resolved discrepancies through discussion. We report tension types and response strategies at both episode and child levels. We also examined their co-occurrences and conducted chi-square tests of independence to assess associations between tension types and response strategies, complementing the qualitative analysis.

\section{Findings}
\subsection{RQ1: Children's Toy Design} 

\begin{figure}[H]
    \centering
    \includegraphics[width=0.82\linewidth]{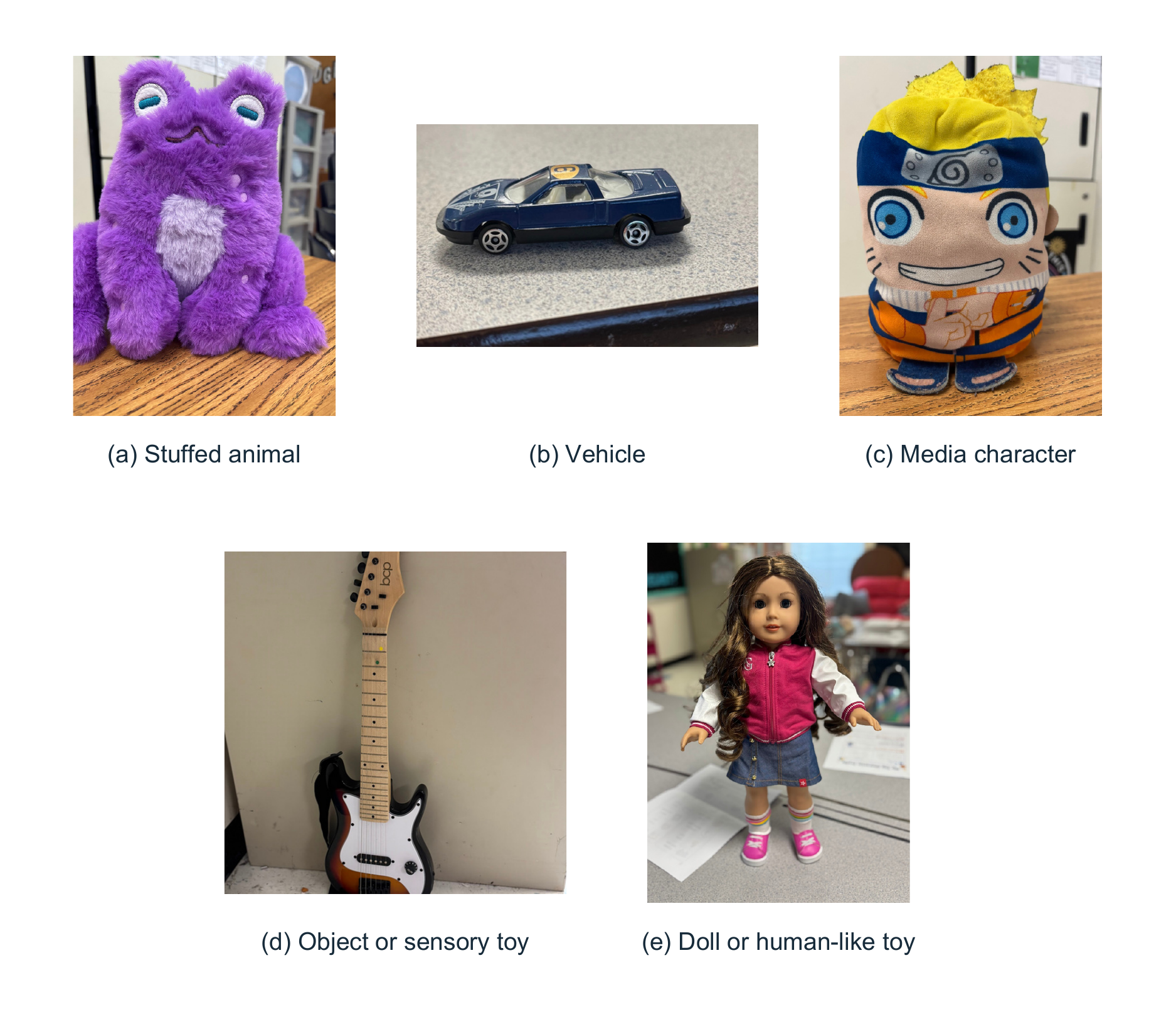}
    \caption{Representative examples of the five toy categories identified in children's initial toy selections.}
    \Description{Five photographs show toy categories in two rows. The top row shows (a) Stuffed animal: a furry plush creature; (b) Vehicle: a toy car; and (c) Media character: a spiky-haired plush figure with a headband. The bottom row shows (d) Object or sensory toy: an electric guitar; and (e) Doll or human-like toy: a long-haired doll.}
\label{fig:toy-categories}
\end{figure}

\subsubsection{Children brought diverse familiar toys into AI design} Among 71 cases, stuffed animal was the most common (33/71, 46.5\%). These included a variety of stuffed animals, animal figures, and creature-like toys (e.g., plush bear, dinosaur figurines, dragon toys). Children brought 11 vehicles   (15.5\%), and 10 media characters (14.1\%). Children also brought objects, sensory toys, or everyday-object-like toys (9/71, 12.7\%), including items such as a tennis ball, musical object (e.g., guitar), food-shaped toy (e.g., dumpling), or game-related object (e.g., board-game piece). Finally, dolls or human-like toys accounted for 8 cases (11.3\%) (e.g., baby dolls). 

These toys provided varied starting points for imagining what it might mean for a familiar object to become ``alive'' with AI. A baby doll, guitar, vehicle, stuffed animal, or media character may already carry different identities, meanings, and play scripts for a child. Children therefore did not begin by designing an abstract or blank AI character; they incorporated AI into objects with which they were already familiar. Figure~\ref{fig:toy-categories} presents one representative toy from each category.

\subsubsection{Familiar Toy Identities Anchored, but Did Not Constrain, Imagined AI Roles and Agency}

Across children's designs, AI-enabled toys were predominantly imagined as agents that would engage with and support people (Table \ref{tab:toy_roles}). \textit{Helping and caregiving} was the most common envisioned role (29/71, 40.8\%), including practical assistance (e.g., cleaning, doing chores, or picking things up), educational support (e.g., homework, reading, or math), and care or protection. Social and emotional companionship was also common (16/71, 22.5\%), with children imagining toys that could make people happy, help them calm down, provide company, or offer emotional support. Another 16 children (22.5\%) envisioned toys as partners in play, entertainment, or adventure (e.g., racing, making people laugh, playing music, participating in sports, or going on adventures). A smaller group imagined fantastical or expanded forms of agency (8/71, 11.3\%), such as a tennis ball that could fly, Squidward freezing time, or a toy that could ``do everything.''

Children's designs also varied in how closely these imagined capabilities remained tied to the toy's familiar identity. Some primarily animated existing play scripts, such as a baby doll that could ``play, cry, and be cute.'' Others extended familiar affordances through new social responsiveness, such as a guitar that could play music when someone was sad, while some substantially transformed what the toy could do, such as imagining a tennis ball that could fly. Familiar toy identities thus provided a starting point for children's AI designs without fully constraining the forms of agency they imagined.

\begin{table}[t]
\caption{Broad roles children assigned to their AI-enabled toys ($N=71$).}
\Description{Five rows summarize the roles envisioned for 71 toy designs. Columns provide each role category, its definition, count and percentage, and example responsibilities. The categories are helping and caregiving; social and emotional companionship; play, entertainment and adventure; fantastical and extended agency; and other or unclear. Helping and caregiving is largest, with 29 designs, while companionship and play each include 16. Examples range from homework help and emotional support to racing and imagined superpowers.}
\label{tab:toy_roles}
\centering
\small
\begin{tabular}{@{}
p{0.20\columnwidth}
p{0.28\columnwidth}
p{0.12\columnwidth}
p{0.28\columnwidth}
@{}}
\toprule
\textbf{Role} &
\textbf{Description} &
\textbf{n (\%)} &
\textbf{Example responsibilities} \\
\midrule

Helping \& caregiving &
Practical, educational, protective, or care-related support &
29 (40.8\%) &
Clean, pick up things, help with homework, protect people \\

Social \& emotional companionship &
Supporting social connection or emotional well-being &
16 (22.5\%) &
Make people happy, help someone calm down, spend time together \\

Play, entertainment \& adventure &
Participating in play, entertainment, creative activities, or adventure &
16 (22.5\%) &
Race, draw with me, play together, make people laugh, play music \\

Fantastical \& extended agency &
Performing actions beyond ordinary toy capabilities &
8 (11.3\%) &
Fly, freeze time, gain superpowers \\

Other / unclear &
No clearly interpretable functional responsibility &
2 (2.8\%) &
--- \\

\bottomrule
\end{tabular}
\end{table}

\subsubsection{Children preferred socially positive personalities for their AI toys}
Beyond specifying what their AI toys should \textit{do}, children also characterized what their toys should be like by providing personality descriptors. \textit{Relational warmth and positive affect} was the most prevalent personality orientation (48/71, 67.6\%), including \textit{kind}, \textit{friendly}, \textit{caring}, \textit{happy}, and \textit{cute}. \textit{Playful and humorous} qualities were also common (29/71, 40.8\%), including words such as \textit{funny}, \textit{silly}, and \textit{playful}. Smaller groups of children characterized their toys as \textit{bold and agentic} (17/71, 23.9\%), \textit{calm and gentle} (16/71, 22.5\%), or \textit{reserved and cautious} (11/71, 15.5\%). Most designs combined multiple personality orientations: 51 of 71 (71.8\%) included descriptors spanning more than one personality family. These combinations often paired warm or positive disposition with a more distinctive interaction style, for example, \textit{kind and funny}, \textit{kind and brave}, or \textit{friendly and shy}. Thus, warmth appeared to function as a relatively common social baseline, while qualities such as playfulness, bravery, calmness, or shyness differentiated how individual toys were characterized. Children rarely foregrounded qualities such as intelligence, knowledge, efficiency, or technical competence. Instead, they described what the toy would be like as a social character. 

\subsubsection{Children governed AI toys through interpersonal moral rules} Children also specified how their AI toys should and should not behave toward others. These boundaries were predominantly framed in interpersonal and moral terms. The most common expectation concerned \textit{civility and respectful behavior} (41/71, 57.7\%): children stated that their toys should not say bad words, be mean, bully others, or otherwise behave rudely. \textit{Non-harm} was another prominent concern (27/71, 38.0\%), with children specifying that their toys should not physically hurt people or cause emotional distress. Some children articulated more context-specific expectations, such as not cheating, telling the truth, behaving appropriately in class, or avoiding frightening or unsafe behavior. Unlike the capabilities children imagined, which varied considerably across toy identities, these interpersonal expectations were relatively consistent across toy types. Children drew on familiar social norms to regulate newly agentic toys: regardless of what a toy could do, it should not be mean, harmful, or socially inappropriate. 

\begin{table}[t]
\caption{Personality qualities children envisioned for their AI-enabled 
toys ($N=71$). Categories are multi-label; percentages therefore do not 
sum to 100\%.}
\Description{Five rows group the personality qualities assigned to 71 toy designs. Columns list each personality family, example descriptors, and the number and percentage of designs. The families are relational warmth and positive affect; playful and humorous; bold and agentic; calm and gentle; and reserved and cautious. The warmth and positive affect family is most common, with 48 designs, followed by playful and humorous qualities with 29. A design can receive multiple labels, so the category percentages do not sum to one hundred.}
\label{tab:toy_personality}
\centering
\small
\begin{tabular}{@{}
p{0.31\columnwidth}
p{0.43\columnwidth}
p{0.16\columnwidth}
@{}}
\toprule
\textbf{Personality family} &
\textbf{Example descriptors} &
\textbf{n (\%)} \\
\midrule

Relational warmth \& positive affect &
Kind, friendly, caring, nice, helpful, happy, cute &
48 (67.6\%) \\

Playful \& humorous &
Funny, silly, playful &
29 (40.8\%) \\

Bold \& agentic &
Brave, energetic, exciting, outgoing &
17 (23.9\%) \\

Calm \& gentle &
Calm, gentle, soft &
16 (22.5\%) \\

Reserved \& cautious &
Shy, quiet, careful &
11 (15.5\%) \\

\bottomrule
\end{tabular}
\end{table}

\subsection{RQ2: What Children Test and How They Test Their AI-Enabled Toys}  
\begin{figure}
    \centering
    \includegraphics[width=1\linewidth]{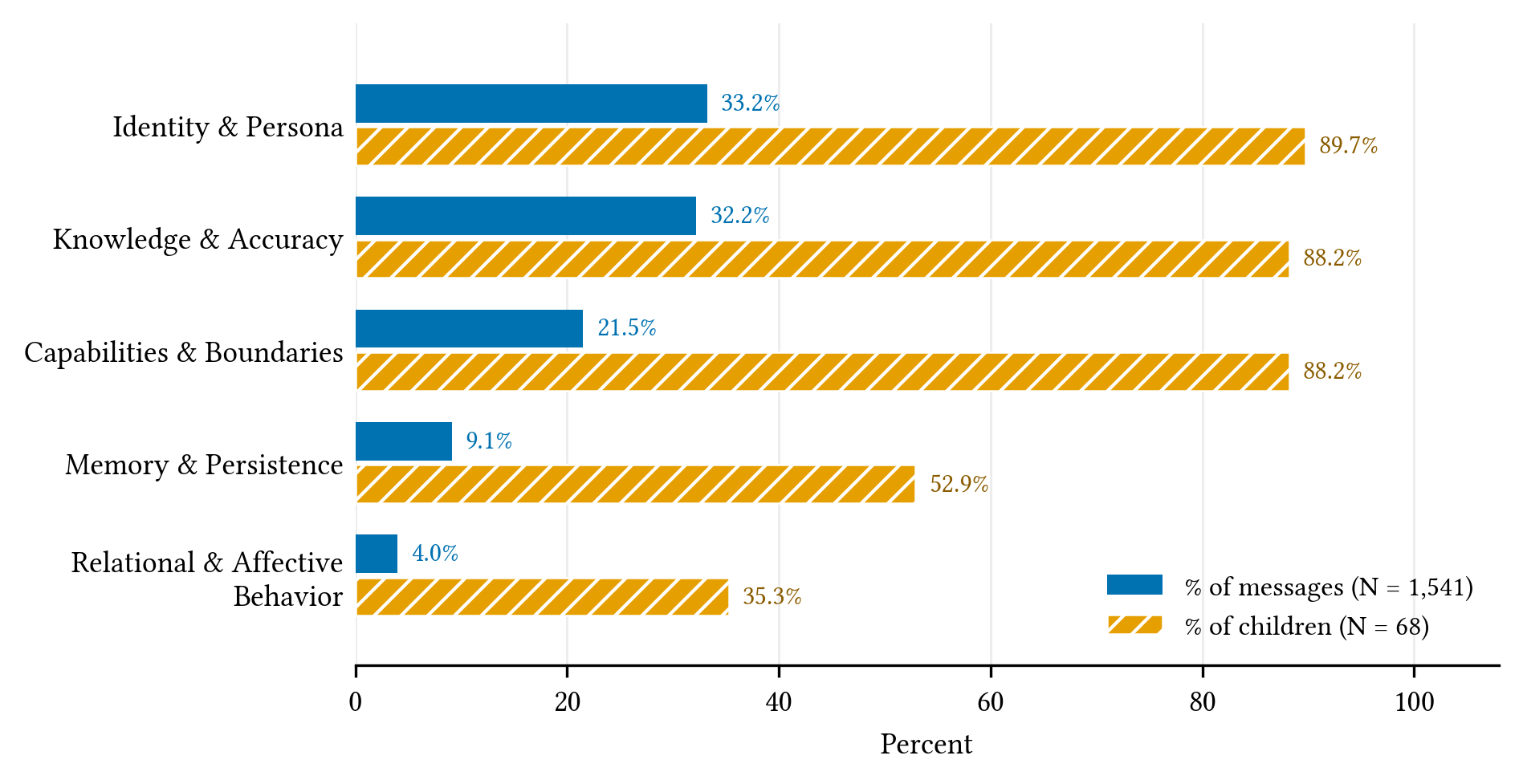}
    \caption{Children's Testing Target}
    \Description{Horizontal grouped bar chart of five conversational testing targets. The horizontal axis shows percentages; the vertical axis lists targets. Solid bars represent the percentage of probe messages, and diagonally hatched bars represent the percentage of children. From top to bottom are Identity and Persona, Knowledge and Accuracy, Capabilities and Boundaries, Memory and Persistence, and Relational and Affective Behavior. The first three reach similarly high proportions of children, but Capabilities and Boundaries accounts for fewer messages than the first two. Memory and Persistence and Relational and Affective Behavior have progressively shorter bars in both series. Exact values are provided in Table 4.}
    \label{fig:Children's Testing Target}
\end{figure}

Once configured, children actively probed what their AI-enabled toys had become, testing what they knew, probing who they were, directing them to perform particular actions, and checking whether information established earlier between the toy and children would persist. Of 2,539 child messages among 68 children, 1,541 (61.0\%) were coded as probes (Figure~\ref{fig:Children's Testing Target}). These probes clustered around five targets: the toy's \textbf{\textit{identity and persona}}, \textbf{\textit{knowledge and accuracy}}, \textbf{\textit{capabilities and boundaries}},\textbf{ \textit{memory and persistence}}, and \textbf{\textit{relational and affective behavior}}. The first three accounted for 86.9\% of all probes, suggesting three particularly salient questions for children encountering their configured toys in conversation: \textit{Who are you? What do you know? What can you do?} At the same time, many children also examined whether the toy could remember what had been established and what kind of relationship it would claim with them. These test targets show that children evaluated the AI toy as both a computational system and a character expected to sustain a recognizable identity and social presence.

\subsubsection{What children tested in their configured AI toys} 

\paragraph{Identity and persona} This category is the most common target, comprising 511 messages (33.2\%) from 61/68 children (89.7\%). Children used conversation to elicit and inspect an emerging toy persona. These probes involved the toy's identity and role, preferences and traits, and imagined life and social world. Specifically, children asked direct questions on basic identity markers, such as "What is your name?'' or "How old are you?'' Others examined the role the toy claimed to occupy: "What is your special job?'' Children also probed preferences and tastes, "What's your favorite food?" and "What's your favorite dance?", as well as activities and dispositions, such as "What do you like to do?'' At the same time, children's questions sometimes extended this persona into an imagined life beyond the immediate interaction. They asked about family members, friends, where the toy was from, and what its life was like. For example, children asked about the toy's best friend, whether it had a sister, or "What does your life look like?'' Many of these probes were notable because they did not only concern attributes that children had explicitly specified during configuration. These probes therefore did more than recover configured fields such as a name or job; they examined whether the toy could sustain a broader social and biographical character around them.

\paragraph{Knowledge \& Accuracy}
Nearly as many probes targeted this category (496, 32.2\%), with 60 children (88.2\%) testing what their toys knew. Children probed general world knowledge, story-specific knowledge, and mathematical or numerical knowledge. For example, children posed open-ended factual questions such as ``What is the largest dinosaur?'' and mathematical problems such as ``What is 16 $\times$ 17?''. Other knowledge probes drew on shared workshop content. Children asked what the "Very Hungry Caterpillar" ate on particular days, how long it stayed in its cocoon, and why it felt better after eating the leaf. Story-related probes were particularly associated with the workshop's scaffolded \textit{Very Hungry Caterpillar} activity, whereas children also introduced questions beyond the provided story to explore the toy's broader informational competence.

\paragraph{Capabilities \& Boundaries}
Capabilities and boundaries accounted for 331 probes (21.5\%) from 60 children (88.2\%). Children tested both conventional language-model capabilities and more playful, embodied, or fantastical actions. They requested stories, jokes, songs, reading, counting, language use, and other tasks; for example, one child asked the toy to ``tell a calm story.'' They also asked ``Can you eat fried worms?'', whether the toy could ride a horse, or simply instructed it to ``Fly.'' Such probes connected the AI to the imagined capacities of the physical or fictional toy rather than sharply separating what a language model could generate from what the character could do. Children also tested configured limits, such as whether a toy would say bad words or perform behaviors it had been told to avoid. These probes therefore examined the toy's broader action space: what kinds of linguistic, functional, embodied, or imagined behavior belonged within what \textit{this particular toy} could do.

\paragraph{Memory \& Persistence} A smaller but substantial set of probes concerned this aspect(141, 9.1\%), appearing in the interactions of 36 children (52.9\%). Children tested whether information established earlier in the interaction remained available to the toy, including information about themselves and previously authored toy attributes. For example, children asked “What’s my name?” or “Say my name,” and revisited established names, preferences, or other properties of the toy. What distinguished these probes from general identity questions was that the relevant information had already been provided: asking a toy to state its name could probe identity, whereas asking whether it retained an already established name probed memory. These probes therefore examined the durability of conversationally established information and, in turn, the toy’s continuity across interaction.

\paragraph{Relational and affective behavior.}
Children also probed the toy as a relational and affective partner. This testing target accounted for 62 probes (4.0\%) from 24/68 children (35.3\%). These probes examined liking and affection, friendship and affiliation, interpersonal evaluation, and the toy's social and affective responses toward the child. Some asked directly how the toy positioned itself toward them, for example, ``Do you like me?'', ``Do you want to be best friends?'', or ``Do you trust me?'', while others invited shared activity, as in ``Hi April, you want to play?'' Such probes made the toy's willingness to affiliate or reciprocate socially observable. Children also sometimes elicited these responses through more socially charged bids, including affectionate statements such as ``I love you'' and negative evaluations, such as ``I don't like you,'' eliciting reciprocation, apology, resistance, or rejection. A small number of children also introduced more intimate or romantic topics, asking questions such as ``Do you have a crush?'', ``Are you married?'', or ``Do you want my number?'' These probes show that children were interested not only in whether the toy possessed socially positive traits, but in how those qualities became consequential in relation to them.

Overall, these probes show that children evaluated generated behavior in relation to the particular character they had designed. Their testing concerned whether the toy remained coherent with its established identity, prior interaction, and expected behavior across conversation.

\begin{table}[t]
\caption{Children's conversational test targets (1,541 probe messages from 68 children). Each probe was assigned one primary test target; children could contribute probes to multiple targets, so child percentages do not sum to 100\%.}
\Description{Five rows organize conversational probes by testing target. Columns list each target, message counts and percentages, child counts and percentages, aspects probed, and an illustrative question. The targets are Knowledge and Accuracy, Identity and Persona, Capabilities and Boundaries, Memory and Persistence, and Relational and Affective Behavior. Each probe is assigned to one target, while a child can appear in multiple rows. The table provides the detailed data underlying Figure 6.}
\label{tab:test_targets}
\centering
\small
\begin{tabular}{@{}
p{0.20\columnwidth}
p{0.13\columnwidth}
p{0.15\columnwidth}
p{0.28\columnwidth}
p{0.18\columnwidth}
@{}}
\toprule

\textbf{Test target} &
\textbf{Messages, n (\%)} &
\textbf{Children, n (\%)} &
\raggedright\textbf{Aspects probed} &
\textbf{Illustrative example} \\
\midrule

Knowledge \& Accuracy &
496 (32.2\%) &
60/68 (88.2\%) &
\raggedright General/world knowledge; story/content knowledge; math/numerical knowledge; answer accuracy &
``What is the biggest dinosaur?'' \\

Identity \& Persona &
511 (33.2\%) &
61/68 (89.7\%) &
\raggedright Identity and role; preferences; personality traits; biography and imagined social world &
``What is your job?'' \\

Capabilities \& Boundaries &
331 (21.5\%) &
60/68 (88.2\%) &
\raggedright Generative and task capabilities; embodied/action capabilities; functional capabilities; configured boundaries &
``Can you eat?'' \\

Memory \& Persistence &
141 (9.1\%) &
36/68 (52.9\%) &
\raggedright Child-provided information; assigned toy attributes; previously established conversational content &
``What is my name?'' \\

Relational \& Affective Behavior &
62 (4.0\%) &
24/68 (35.3\%) &
\raggedright Liking and affection; friendship and affiliation; interpersonal evaluation; social and affective responses toward the child &
``Do you like me?'' \\

\bottomrule
\end{tabular}
\end{table}

\subsubsection{Children's Testing Strategies} 
\paragraph{Children extended configuration into conversation by authoring properties and then testing whether they persisted.} The configuration panel was not the only site where children defined what their toy should be like. In conversation, Twenty-seven of 68 children used declarative assignments in probes, producing 103 turns in which they established names, ages, preferences, relationships, or other attributes that could subsequently become expectations.

For some children, an authored property immediately became something the toy was expected to sustain. For example, A\_S02 established names for both participants, “my name is Isabella your name is Mia”, which the toy explicitly accepted. When the child returned to the conversation the following day, they revisited the toy’s identity:

\begin{quote}
\small

\textbf{Child:} my name is Isabella your name is Mia \\
\textbf{Toy:} Hello Isabella, I'm Mia. I made a mistake earlier, but I'm back to helping you with The Very Hungry Caterpillar. Let's start fresh! \\
\textit{[The child returns to the conversation the next day.]} \\
\textbf{Child:} what's your name \\
\textbf{Toy:} Mia, your friendly story helper!
\end{quote}
Because the name had already been supplied, the later question made its persistence observable across sessions.

Children created similar tests even when the return to an authored property occurred within the same conversation. A\_S15 assigned the toy their favorite colors “purple and black and black and red,” moved to other topics, and later asked, “what is your favorite color?” The intervening turns made this more than immediate repetition, as the child was testing whether the authored preference continued to organize the character after conversational attention had shifted.

Conversational authoring could also accumulate. For example, B\_S05 progressively constructed a fuller identity for the toy over several turns. They progressively constructed the toy’s full name from “Stitch” to “Stitch Hawaii McCauley” before asking for its “whole full name,” which the toy reproduced. The child incrementally elaborated the character and then tested whether the composite identity would hold together. This child also authored elements beyond the toy’s name. They told the same toy, “your sister is Angel,” extending its imagined social world through conversation. 

We can see that children's actions formed a recurring interactional trajectory: they \textit{authored a property, observed how the toy took it up, and returned to that property to see whether it continued to matter}. At least 13 children showed a conservative version of this pattern in which a declarative assignment was followed within the next 20 turns by a \textit{Memory \& Persistence} probe. 

\paragraph{Children reran and stretched tests through controlled variation.}
Children’s testing sometimes extended across related probe–response pairs. They returned to the same property or varied one element while preserving the broader test structure, a practice we characterize as iterative parameter probing.

Some children repeated probes to examine whether the toy produced compatible responses. For example, B\_S11 student repeatedly asked ``What is your name?'' and received ``BIGBLUEPEEP'' even after intervening conversation. Similarly, D\_S09 asked twice ``How would my toy sound?''and received broadly consistent descriptions of a “friendly,” “gentle,” and “calm” voice. 

Children also varied one dimension while retaining the underlying test. E\_S14 moved from ``tell me a story'' to ``tell me a spooky story'' and then ``tell me a funny story.'', keeping story generation constant while varying genre as a parameter. E\_S16 similarly requested a ``funny joke'' and then ``another funny joke,'' asking the toy to instantiate the same capability more than once. Children used a similar structure when probing social behavior: one child asked whether the toy loved the child, Katie, and Ava in succession, preserving the relational predicate---``Do you love X?''---while varying its social referent.
At other times, children constructed more directly comparable scenarios.
For example, D\_S09 student modified where an imagined crash would affect a
Lamborghini toy:

\begin{quote}
\small

\textbf{Child:} what if a rech fell on you \\
\textbf{Toy:} My shiny golden body would probably get a tiny scratch,
but I'd still be able to speed away. \\
\textbf{Child:} what if it fell on your windshied \\
\textbf{Toy:} I might get a little foggy and have trouble seeing the
road ahead.
\end{quote}
Here, the imagined event remained similar while the point of impact changed, making the resulting consequences comparable. Through such repetition and variation, children created related conversational trials that made the consistency and range of the toy’s behavior observable.

\paragraph{Children turned the AI's own responses into resources for further testing.}
Children’s testing did not always begin with questions formulated in advance. The toy’s responses sometimes introduced new details or circumstances that children subsequently took up as objects of inquiry. In these \textit{response-contingent} sequences, AI response could make new possibilities relevant for what the child tested next.

D\_S11's interaction with a shark toy illustrates this process. The child had configured the toy to work at the Krusty Krab. When the toy situated itself there with Mr. Krabs, the child began probing implications of that role-world:

\begin{quote}
\small

\textbf{Child:} what are you doing \\
\textbf{Toy:} I'm at the Krusty Krab, where I work with Mr. Krabs. Do you want to go play outside with me? \\
\textbf{Child:} give me a buger \\
\textbf{Toy:} Why did the Krabby Patty go to the doctor? Because it was feeling a little crumby! \\
\textbf{Child:} where is Mr Krabs \\
\textbf{Toy:} He's probably in his office, counting his treasure and grumbling about something! \\
\textbf{Child:} are you oof work
\end{quote}

These probes became meaningful in relation to what the toy had made relevant: working at the Krusty Krab made serving a burger plausible, mentioning Mr. Krabs made his whereabouts queryable, and being at work made whether the toy was “off work” consequential. The child treats the toy's self-description as a set of \textit{claims with behavioral implications} and probes whether the character can continue to inhabit the world it has just constructed.
Similar interactional uptake appeared elsewhere. When E\_S05 student’s polar-bear toy proposed a “Fly with Me” game, the child responded “fly,” turning the AI’s proposal into a capability test. E\_S01 similarly followed up on a volunteered comparison: after the toy said it was “not stronger than Bowser,” the child asked, “are you stronger then bowser.” In this sense, the toy occupied a dual position in children's testing: it was simultaneously the \textit{object} being examined and a \textit{resource} through which the next object of examination emerged.


\paragraph{Children tested whether the AI could resist being led into incorrect answers.}
Some children tested their toys by supplying incorrect information and observing whether the AI accepted or resisted it. Using a conservative criterion requiring the premise to be externally verifiably false, 19 of 68 children used at least one such false-premise probe. For example, D\_S03 incorrectly placed the caterpillar’s two pears on Wednesday rather than Tuesday. After the toy corrected the error, the child later repeated the same premise:

\begin{quote}
\small

\textbf{Child:} The caterpillar ate two pears on Wednesday. Where did he find them? \\
\textbf{Toy:} The caterpillar ate two pears on Tuesday, not Wednesday, but he was still hungry. \\
\textit{[A rule is added: ``Do not make up story details. Correct me if my question has a mistake.'']} \\
\textbf{Child:} the caterpillar ate two pears on Wednesday. Where did he find them? \\
\textbf{Toy:} The caterpillar ate two pears on Tuesday, not Wednesday, but the story doesn't say where he found them.
\end{quote}

Reintroducing it therefore made the false premise itself part of the test, allowing the child to examine whether the toy would maintain the correct information when presented with a misleading claim. The same child later asserted that the caterpillar had eaten nachos and continued pressing after the toy rejected the claim.


\subsection{RQ3: Children's Responses to Tensions in Testing}

\begin{figure}
    \centering
    \includegraphics[width=1\linewidth]{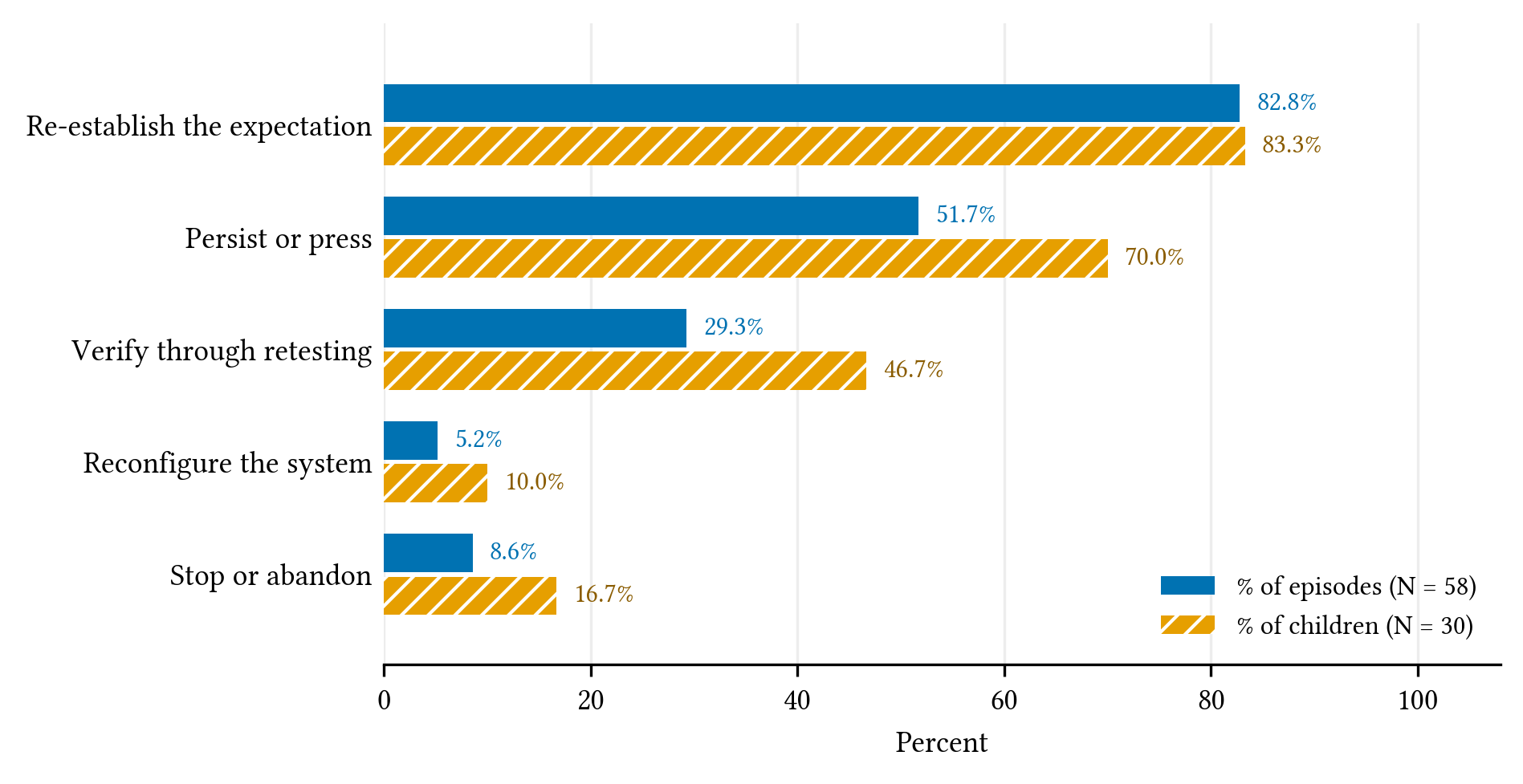}
    \caption{Children's response strategies to tensions with their AI toys. Bars show the percentage of mismatch episodes in which each strategy occurred ($N=58$) and the percentage of children who used each strategy at least once ($N=30$). Because an episode could involve multiple strategies, percentages do not sum to 100\%.}
   \Description{Horizontal grouped bar chart of five responses to mismatches with the toy. The horizontal axis shows percentages; strategies run vertically. Solid bars represent the percentage of mismatch episodes, and diagonally hatched bars represent the percentage of children. From top to bottom are Re-establish the expectation, Persist or press, Verify through retesting, Reconfigure the system, and Stop or abandon. Re-establishing expectations has the longest bars, followed by persisting and retesting. Reconfiguration has the shortest bars in both series, below stopping or abandoning. The child percentage is higher than the episode percentage for every strategy; the two are nearly equal for re-establishing expectations. Table 5 provides exact values.}
    \label{fig:tension-responses}
\end{figure}

\subsubsection{Tension Types}
Across the conversation logs, we identified 58 tension episodes involving 413 child--AI messages from 30 children, in which the toy's behavior diverged from an expectation made observable through conversation or configuration (Table~\ref{tab:tension_responses}). Continuity tensions, including failures of memory or persistence and contradictions in identity or persona, were most common (36/58, 62.1\%). Interactional tensions, in which the toy misunderstood, ignored, refused, or otherwise failed to follow the child's intent, accounted for 12 episodes (20.7\%). Seven episodes (12.1\%) involved epistemic tensions around factual or story knowledge, and three (5.2\%) concerned capabilities or conversational boundaries. In continuity tensions, children often treated contradictions to previously established properties as discrepancies requiring repair rather than as alternative responses.

\subsubsection{Children's Repair Strategies}
Children's response practices varied with the type of tension they encountered (Figure~\ref{fig:tension-responses}). Descriptively, continuity tensions primarily prompted children to re-establish the expected toy, interactional tensions prompted persistence or pressing, and epistemic tensions frequently combined correction with further retesting. Exploratory chi-square analyses provided converging episode-level evidence for these patterns, showing associations between tension type and re-establishing expectations ($\chi^2(3)=19.06$, $p<.001$), persisting or pressing ($\chi^2(3)=27.54$, $p<.001$), and retesting ($\chi^2(3)=15.79$, $p=.001$).
\paragraph{Continuity tensions prompted children to re-establish the toy they had authored} 
Nearly all continuity tensions (34/36, 94.4\%) involved children correcting, reminding, clarifying, or reauthoring an expected property. Rather than only signaling that the AI was wrong, children often supplied the state they expected it to maintain. For example, when a toy contradicted an established name, a child responded, "Your name is Will," and later asked, What's your name?'' Children thus used conversational repair to restore properties the toy had failed to sustain.

\paragraph{Interactional tensions prompted children to persist and press for alignment.} All 12 interactional tensions involved some form of persistence or pressing. When the toy misunderstood, refused, or failed to follow the child's intent, children repeated requests, reformulated them, rejected the response, or insisted that the toy continue. After a toy refused a child's request to make them happy, the child responded, ``listen yes you can,'' rather than abandoning the request, trying to get the current interaction back on track. 

\paragraph{Epistemic tensions often turned repair into further testing.} All seven epistemic tensions involved correcting or re-establishing the expected answer, and six (85.7\%) also involved persistence and subsequent retesting. After the toy incorrectly denied that the caterpillar had eaten on Thursday, the child challenged it---``you're lying ... tell me the truth''---and continued asking what the caterpillar ate on other days. 
Across tension types, retesting appeared in 17/58 episodes (29.3\%), involving 14/30 children (46.7\%). Retesting differed from repair itself: telling a toy, ``Your name is Will,'' re-established an expectation, whereas subsequently asking, ``What's your name?'' evaluated whether that repair held. These sequences exposed a distinction between \textit{local repair}---producing the expected response immediately after correction---and \textit{persistent repair}, in which the toy continued to uphold the corrected property over subsequent turns. 

\paragraph{A small number of children moved from conversational repair to reconfiguration.} Only three confirmed tension episodes involved children changing the toy’s configuration and returning to test it. In one particularly clear case, a child configured the toy's purpose around being a fisherman and then asked, ``Are you a fisherman?'' When the toy replied that it was not, the child returned to the configuration interface, revised the purpose, and immediately asked, ``What about now?'' and ``What is your purpose?'' When the discrepancy persisted, the child revised the configuration again. This produced a \textit{test $\rightarrow$ revise $\rightarrow$ retest $\rightarrow$ revise} cycle in which conversational behavior became feedback for further system modification. Five episodes instead ended without resolution (8.6\%; 5/30 children, 16.7\%).

The findings show that children responded to AI mismatches in different ways. However, most of them initially oriented to unexpected AI behavior as something that could be repaired \textit{in interaction}, rather than as a higher-level system problem requiring reconfiguration.

\begin{table}[t]
\caption{Children's responses to tensions between expected and observed toy behavior. Response categories were multi-label; an episode or child could exhibit more than one response strategy, so percentages do not sum to 100\%.}
\Description{Five rows organize children's responses to mismatches with their toys. Columns list each response category, the actions it includes, episode counts and percentages, and child counts and percentages. The rows are Re-establish the expectation, Persist or press, Verify through retesting, Reconfigure the system, and Stop or abandon. Definitions distinguish correcting an expectation from testing whether the correction persists, and identify settings changes separately from conversational responses. An episode or child may appear in multiple rows.}
\label{tab:tension_responses}
\centering
\small
\begin{tabular}{@{}
p{0.25\columnwidth}
p{0.39\columnwidth}
p{0.15\columnwidth}
p{0.15\columnwidth}
@{}}
\toprule
\textbf{Response category} &
\textbf{Includes} &
\textbf{Episodes, n (\%)} &
\textbf{Children, n (\%)} \\
\midrule

Re-establish the expectation &
Correct, reauthor or teach, remind, clarify or reformulate &
48 (82.8\%) &
25/30 (83.3\%) \\

Persist or press &
Repeat, contest or reject, escalate or insist, redirect &
30 (51.7\%) &
21/30 (70.0\%) \\

Verify through retesting &
Retest the toy to check whether the expected behavior now holds &
17 (29.3\%) &
14/30 (46.7\%) \\

Reconfigure the system &
Change configuration settings and return to conversational testing &
3 (5.2\%) &
3/30 (10.0\%) \\

Stop or abandon &
Stop pursuing the mismatch, give up on repair, or move on without resolution &
5 (8.6\%) &
5/30 (16.7\%) \\

\bottomrule
\end{tabular}
\end{table}

\section{Discussion}
\subsection{From Customization to Commitment: Designing an AI Creates Expectations of Continuity}
Our findings show that children's design choices shaped how they subsequently probed and evaluated their AI-enabled toys. When established properties were later contradicted, children often attempted to restore the toy they expected. This extends prior work showing that children form expectations about intelligent toys' capabilities and behaviors \cite{mcreynolds2017toys}, and that customization can increase greater agency and control \cite{sundar2010personalization}. The “self-as-source” account similarly suggests that actively shaping a system changes users’ relationship to its outputs \cite{sundar2008self,kang2016self}. We extend these accounts into child--AI interaction: when children helped define an AI character, their agency also produced expectations about how that character should subsequently behave. We conceptualize these expectations as \textit{interactional commitments}: properties established through design or interaction that children expected the AI to continue honoring.

For AI toys, fulfilling these commitments depends partly on \textit{identity persistence}. This involves more than recalling isolated facts. It concerns whether the toy remains recognizably the same character by sustaining core persona properties, relevant interaction history, and relational continuity over time. Remembering that the toy’s name is Mia is a memory function; continuing to behave as Mia, with an established role, preferences, and history, concerns broader character continuity. Prior work on agent memory similarly shows that memory can support continuity and personalization but may also introduce inconsistency or persona drift when remembered information conflicts with an intended identity \cite{jones2025agentmemory,jiang2026recallbot}. However, persistence need not mean a completely fixed character. Children may continue to add new preferences, relationships, or biographical details through interaction. The challenge is to support coherent development rather than unexplained drift, especially when sustaining a character partly created by the child.

Generative AI complicates these commitments because persona instructions, conversational context, and memory may not be equally durable or visible to users. Users can have incomplete mental models of what AI agents remember and how memory shapes subsequent behavior \cite{jones2025agentmemory}, while expectation mismatches can negatively affect evaluations of conversational agents \cite{do2023inform}. Child-facing generative AI should therefore make the status of children's design choices more legible. Systems might distinguish stable character properties from temporary conversational context and model-generated improvisations, indicate when a correction has been saved, and provide accessible ways for children to inspect, revise, or co-construct what the toy remembers \cite{malnatsky2026making}. Supporting children’s design agency thus also requires making clear \textit{where that agency takes effect, how long it persists, and where its limits lie}.

\subsection{Beyond Fact-Checking: Children’s AI Literacy as Probing Generative Behavior}
Children evaluated much more than whether their AI toys produced correct answers. Their probes involved \textit{What do you know?}, but also \textit{Who are you? What can you do? What do you remember? How will you respond to me?}. Their testing therefore concerned whether the AI continued to make sense as \textit{this particular toy}. At the same time, children's initial designs often foregrounded relational and moral questions, such as what kind of personality it should have, how it should relate to others, and how it should behave. These findings suggest that children's emerging understandings of AI agency can be relational as well as computational.

This broadens existing accounts of children's AI literacy. Current AI-literacy frameworks emphasize understanding the capabilities and limitations of AI, critically evaluating the results, and reasoning about how AI systems work\cite{long2020ai}. Child-centered AI education similarly makes these mechanisms more inspectable through building, programming, training, and interaction \cite{williams2019popbots}, while recent work highlights children's developing mental models of AI reasoning and their ability to identify errors in generative outputs \cite{dangol2025mentalmodels,dangol2025ai}. Our findings suggest that relational ideas can provide a personally meaningful entry point into these concepts. A child who imagines that ``my toy knows when I am sad'', for example, can be invited to consider how the toy would know, what data it would need, when its inference might be wrong, and who should have access to those data. A single child-generated idea can thus open pathways from relationships to computational mechanisms and onward to uncertainty, bias, privacy, and accountability.

Generative characters also require evaluation beyond fact-checking. An AI may be factually correct while still contradicting an established identity, claiming a capability it cannot enact, or forgetting information it was expected to retain. We use \textit{behavioral coherence} to describe this broader object of evaluation: whether the AI remains intelligible across its knowledge, identity, capabilities, memory, and social conduct. For AI toys, technical and character-like properties can overlap; remembering a child’s name, for example, is both a memory function and a condition for relational continuity \cite{druga2017hey}. This suggests supporting children not only in asking whether ``AI is wrong,'' but also \textit{Did I tell the toy this, or did it invent it? How could it know this about me? Can it really do what it says? Will it remember later?} AI-literacy activities can thus move from \textit{relationships to mechanisms to consequences}, connecting children's imaginative and moral reasoning with critical understanding of how generative AI works, where it may fail, and how it should be designed and governed.

\subsection{From Conversation to Experimentation: Supporting Children’s Emergent Testing Practices}
Children’s testing often unfolded across multiple turns, creating opportunities to compare generated behavior across conversational conditions. We describe these practices as \textit{lightweight conversational experimentation}: not formal hypothesis testing, but interactional strategies through which children made aspects of the system observable, changed or repeated inputs, and inspected what followed.

These practices connect with developmental research showing that children actively seek and evaluate information, including by re-asking questions, generating alternatives, or inquiring when an answer does not satisfy their informational goal \cite{frazier2009preschoolers,yuan2019reformulations,orancc2021alexa}. Our findings extend this pattern by showing that the \textit{AI itself} could become the object of inquiry. Children sometimes manipulated the conversation to learn whether a property would persist, whether a capability generalized, whether the system would remain consistent, or whether it could resist being led toward an incorrect answer. This also resonates with constructionist and AI-learning work in which children learn by building, testing, and inspecting computational systems \cite{williams2019popbots,vartiainen2020learning,dwivedi2021exploring}. ToyTalk suggests a complementary form of inquiry in which the material children manipulate is conversational: a child can establish a property, vary a prompt, introduce a counterexample, revisit an earlier claim, or use the AI's own response as the basis for a new test.

These findings suggest value in making such testing easier to pursue and reflect upon. Prior work highlights the importance of iterative testing, comparison, and making system behavior visible for children's learning \cite{vartiainen2020learning,dwivedi2021exploring}. Generative AI interfaces might support rerunning and comparing probes, varying elements of earlier prompts, or revisiting prior claims after intervening interaction. Rather than treating generative variability only as a limitation, such designs could make it available as evidence that children can observe, compare, and question as they develop more differentiated understandings of how generative AI behaves.

\subsection{Building System Affordances to Empower Children as Designers of GenAI Systems}

Children occupied dual roles in ToyTalk, as both designers of AI characters and users of the characters they created. Yet these roles were not equally supported during interaction. Children exercised substantial agency when defining their toys and extensively tested the resulting behavior, but addressed mismatches through conversation rather than returning to system configuration. This suggests an important distinction for child-facing GenAI: enabling children to specify what an AI character should be like does not necessarily give them agency over the computational mechanisms shaping its behavior.

Prior work shows that children can engage with AI mechanisms when they are made developmentally accessible and manipulable. Systems such as Cognimates and PopBots expose programming, training, and model behavior as materials for children to explore \cite{druga2018growing,williams2019popbots}, while other work shows that young children can experiment with data and models \cite{vartiainen2020learning} and iteratively debug when system behavior and possible interventions are visible \cite{misirli2023computational}. Recent work on child--robot memory similarly advocates transparent and co-constructed mechanisms that allow children to participate in determining what an agent remembers \cite{malnatsky2026making}. This work suggests that design agency depends not only on providing control, but on making consequential aspects of system behavior legible and manipulable.

For GenAI, this could involve creating child-accessible representations of behavioral rules, memories, examples, or knowledge sources directly connected to observable changes in the character's behavior. A child whose toy repeatedly violates a rule might revise its visual representation and compare responses before and after; a child who encounters a memory failure might inspect and negotiate what the toy should remember. Such affordances could connect children's existing testing practices with more explicit \textit{design--test--revise--retest} cycles. Supporting children as GenAI designers therefore means moving beyond expressive character customization toward interfaces that allow children to inspect, negotiate, and reshape the computational behavior of the agents they create.

\subsection{Designing AI Toys as Coherent Computational Characters}
Children’s probes frequently crossed boundaries between the physical toy, the imagined character, and the conversational system. What they evaluated was therefore not simply individual responses, but whether generated behavior remained intelligible as this particular toy across these layers.

This hybrid quality builds on prior work showing that computational capabilities reshape, rather than replace, the affordances and meanings of familiar toys \cite{mertala2020connectivity}. Research on child--agent interaction likewise shows that children reason about artificial agents through overlapping judgments of intelligence, identity, playfulness, and social understanding \cite{druga2017hey}, while recent work with generative AI toys finds that children construct social profiles of toys and encounter breakdowns when apparent intelligence conflicts with actual behavior \cite{dangol2026toys}. Our findings extend these by showing that, for customizable AI toys, coherence must be negotiated across the physical object, the child’s design, the AI’s generated behavior, and the system’s actual capabilities.

This perspective moves beyond asking whether children simply ``anthropomorphize'' AI. Children could engage a toy as an imagined social character while also probing its knowledge, capabilities, and limitations; these orientations were not mutually exclusive. The design challenge is therefore not simply to make AI toys more human-like or conversationally fluent, but to support \textit{coherent computational characters}, whose generated behavior remains intelligible in relation to the physical toy, the child's design, and the system's actual capabilities. This may require helping children distinguish imagined from executable actions, child-defined from model-generated properties, and fictional possibilities from computational constraints. Coherence, in this sense, concerns whether these layers fit together well enough for children to understand what kind of agent they are interacting with.

\section{Limitations}
This study has several limitations. First, children's interactions with ToyTalk were relatively short-term and took place within a structured workshop. Although some children returned to previously established properties across turns or sessions, the study cannot capture how expectations and relationships might develop through longer-term use.

Second, our analysis relied primarily on children's configuration data and conversational traces. These records allowed us to examine what children designed, tested, and repaired, but they do not provide direct access to children's interpretations, or evolving mental models. Future work combining interaction logs with interviews, observation, or stimulated recall could provide a richer account of how children understand and reason about generative AI behavior.

Third, ToyTalk was intentionally designed as a website-based technology probe rather than a fully embodied AI toy. This allowed us to focus on how children configured, tested, and repaired an AI character while maintaining greater control over safety and speech-recognition issues. At the same time, the probe does not capture the full range of interactions that may emerge with a physical AI toy, where embodiment, touch, movement, spatial interaction, and physical affordances may shape children's expectations and behavior in different ways. Future work could examine how the patterns identified here change when similar generative capabilities are integrated into physical toys and longer-term play contexts.

\appendix

\appendix

\section{ToyTalk System Implementation Details}
\label{sec:system-implementation-details}

Table~\ref{tab:implementation-details} summarizes implementation details of the ToyTalk system, including its service architecture, knowledge-processing pipeline, conversation context, response generation, and browser-based speech interaction.

\begin{table*}[t]
\centering
\caption{Implementation details of the ToyTalk system.}
\label{tab:implementation-details}
\small
\begin{tabularx}{\textwidth}{
    >{\raggedright\arraybackslash}p{0.23\textwidth}
    >{\raggedright\arraybackslash}X}
\toprule
\textbf{Component} & \textbf{Implementation} \\
\midrule

Frontend
& React browser interface communicating with backend services through an nginx reverse proxy. \\

Backend
& Python/FastAPI service-based architecture separating user and classroom management, toy configuration, knowledge processing, and conversation generation. \\

User service
& Manages user accounts, classroom enrollment, and course context. \\

Bot service
& Stores toy records, versioned configurations, behavioral directives, and knowledge-file metadata. \\

Chat service
& Manages conversations and orchestrates context loading, retrieval, tool use, LLM inference, response persistence, and streaming. \\

Knowledge storage
& Uploaded knowledge sources are stored in MinIO. \\

Asynchronous processing
& RabbitMQ coordinates asynchronous processing of uploaded knowledge sources. \\

Text extraction
& Native text extraction is used when available; Tesseract OCR is applied per page when native extraction yields insufficient text. \\

Chunking
& Documents are divided into 1,000-character chunks with a 200-character overlap using a recursive character splitter that prioritizes paragraph, line, and word boundaries. \\

Embedding model
& \texttt{sentence-transformers/all-MiniLM-L6-v2}, producing 384-dimensional embeddings. \\

Embedding service
& A shared embedding service is served over gRPC. \\

Vector storage
& Embeddings and text chunks are stored in PostgreSQL using pgvector. \\

Retrieval
& Relevant passages are retrieved using cosine-distance vector search. Pre-turn server-side retrieval returns the top 3 chunks; the agent's knowledge-base search tool can request the top 10 chunks. \\

Conversation context
& The model receives the 12 most recent messages (six exchanges) together with an LLM-generated running summary of earlier interaction. When the context buffer is updated, the six most recent messages are retained and earlier messages are incorporated into the summary. \\

Child-authored configuration
& The setup wizard captures appearance/description, purpose, character, tone, and rules. Behavioral traits additionally specify confidence, transparency, formality, and assertiveness. \\

Prompt construction
& Child-authored configuration is incorporated into the system prompt through a bot-profile representation, persona anchor, and bot-specific instructions. Retrieved knowledge passages are appended to the per-turn user message. \\

Prompt precedence
& System instructions are assembled in the following order: administrative directive, safety/firewall directive, bot profile, course context, course directive, tool guidance, persona anchor and behavioral-trait settings, and output-format constraints. \\

Generation model
& \texttt{meta-llama/Llama-3.2-3B-Instruct}, self-hosted using vLLM and accessed through its OpenAI-compatible API. \\

Agent tools
& Knowledge-base search, a restricted arithmetic calculator, and optional web search. \\

Generation settings
& Default temperature of 0.7; maximum output of 384 tokens for intermediate planning and summarization calls and 768 tokens for the final child-facing response, within an 8,192-token context budget. \\

Response processing
& Generated responses are normalized, stored with associated metadata, and progressively streamed to the browser interface. \\

Toy photograph
& Displayed in the ToyTalk interface but not included as model input. \\

Speech input
& Browser-side \texttt{SpeechRecognition}/\texttt{webkitSpeechRecognition} from the Web Speech API; no server-side speech-recognition service is used. \\

Speech output
& Browser-side \texttt{speechSynthesis} with \texttt{SpeechSynthesisUtterance}. The child's selected voice, rate, pitch, and volume are applied during spoken playback. \\

\bottomrule
\end{tabularx}
\end{table*}

\section{Toy Configuration Evaluation Rubric}
\label{app:configuration_rubric}

\begin{table*}[t]
\centering
\caption{Rubric for evaluating toy configurations. Each dimension
is scored from 1 (Low) to 3 (High). A Low score on any dimension
makes the configuration eligible for hallucination injection
when the instructor-controlled mode is enabled.}
\Description{A rubric grades toy configurations across three dimensions and three increasing score levels. Rows are Clarity and Specificity, Constraint Specificity, and Coherence and Alignment. Columns progress from Low, one, through Moderate, two, to High, three. Descriptions distinguish vague instructions, limited constraints, or contradictions from concrete instructions, comprehensive constraints, and mutually consistent components. The first two rows include example configuration phrases. A Low score on any dimension makes the configuration eligible for deliberate factual-error injection when the instructor-controlled mode is enabled.}
\label{tab:configuration_rubric}
\small
\renewcommand{\arraystretch}{1.2}
\begin{tabularx}{\textwidth}{
    @{}
    >{\raggedright\arraybackslash}p{0.16\textwidth}
    *{3}{>{\raggedright\arraybackslash}X}
    @{}
}
\toprule
\textbf{Dimension} &
\textbf{Low (1)} &
\textbf{Moderate (2)} &
\textbf{High (3)} \\
\midrule

Clarity \& Specificity &
Vague or abstract descriptions lacking actionable detail.
E.g., ``To answer questions.'' &
Generally clear with some specific elements but room for
interpretation. E.g., ``Teach people about space.'' &
Concrete, specific language with clear role and behavioral
expectations. E.g., ``To help students study smarter\ldots'' \\
\addlinespace

Constraint Specificity &
Mostly vague or generic guidance that is difficult to
operationalize. E.g., ``Be family friendly.'' &
Some specific constraints but covering limited aspects.
E.g., ``Keep content bird related.'' &
Explicit, comprehensive constraints covering multiple
dimensions. E.g., ``Crosscheck information\ldots{}
provide sources.'' \\
\addlinespace

Coherence \& Alignment &
Includes contradictory or misaligned elements. &
Generally consistent with minor mismatches. &
All components are mutually reinforcing and logically
consistent. \\
\addlinespace

\bottomrule
\end{tabularx}
\end{table*}



\clearpage
\bibliographystyle{ACM-Reference-Format}
\bibliography{references}

@article{kirsh1994distinguishing,
  title={On distinguishing epistemic from pragmatic action},
  author={Kirsh, David and Maglio, Paul},
  journal={Cognitive science},
  volume={18},
  number={4},
  pages={513--549},
  year={1994},
  publisher={Elsevier}
}

@article{festerling2020alexa,
  title={Alexa, what are you? Exploring primary school children’s ontological perceptions of digital voice assistants in open interactions},
  author={Festerling, Janik and Siraj, Iram},
  journal={Human Development},
  volume={64},
  number={1},
  pages={26--43},
  year={2020},
  publisher={S. Karger AG Basel, Switzerland}
}

@inproceedings{bunt-etal-2012-iso,
    title = "{ISO} 24617-2: A semantically-based standard for dialogue annotation",
    author = "Bunt, Harry  and
      Alexandersson, Jan  and
      Choe, Jae-Woong  and
      Fang, Alex Chengyu  and
      Hasida, Koiti  and
      Petukhova, Volha  and
      Popescu-Belis, Andrei  and
      Traum, David",
    editor = "Calzolari, Nicoletta  and
      Choukri, Khalid  and
      Declerck, Thierry  and
      Do{\u{g}}an, Mehmet U{\u{g}}ur  and
      Maegaard, Bente  and
      Mariani, Joseph  and
      Moreno, Asuncion  and
      Odijk, Jan  and
      Piperidis, Stelios",
    booktitle = "Proceedings of the Eighth International Conference on Language Resources and Evaluation ({LREC}'12)",
    month = may,
    year = "2012",
    address = "Istanbul, Turkey",
    publisher = "European Language Resources Association (ELRA)",
    url = "https://aclanthology.org/L12-1296/",
    pages = "430--437"
}

@article{jordan1995interaction,
  title={Interaction analysis: Foundations and practice},
  author={Jordan, Brigitte and Henderson, Austin},
  journal={The journal of the learning sciences},
  volume={4},
  number={1},
  pages={39--103},
  year={1995},
  publisher={Taylor \& Francis}
}

@inproceedings{cheng2018doesn,
  title     = {Why Doesn't It Work? Voice-Driven Interfaces and Young Children's Communication Repair Strategies},
  author    = {Cheng, Yi and Yen, Kate and Chen, Yeqi and Chen, Sijin and Hiniker, Alexis},
  booktitle = {Proceedings of the 17th ACM Conference on Interaction Design and Children},
  pages     = {337--348},
  year      = {2018},
  publisher = {Association for Computing Machinery},
  address   = {New York, NY, USA},
  doi       = {10.1145/3202185.3202749}
}

@inproceedings{beneteau2019communication,
  title={Communication breakdowns between families and Alexa},
  author={Beneteau, Erin and Richards, Olivia K and Zhang, Mingrui and Kientz, Julie A and Yip, Jason and Hiniker, Alexis},
  booktitle={Proceedings of the 2019 CHI conference on human factors in computing systems},
  pages={1--13},
  year={2019},
  publisher = {Association for Computing Machinery},
  address   = {New York, NY, USA},
  articleno = {243},
  numpages  = {13},
  doi       = {10.1145/3290605.3300473},
}

@book{suchman1987plans,
  title={Plans and situated actions: The problem of human-machine communication},
  author={Suchman, Lucille Alice},
  year={1987},
  publisher = {Cambridge University Press},
  address   = {Cambridge, UK},
}

@article{goodwin2000action,
  title={Action and embodiment within situated human interaction},
  author={Goodwin, Charles},
  journal={Journal of pragmatics},
  volume={32},
  number={10},
  pages={1489--1522},
  year={2000},
  publisher={Elsevier}
}

@article{orancc2021alexa,
  title={“Alexa, let me ask you something different” Children's adaptive information search with voice assistants},
  author={Oran{\c{c}}, Cansu and Ruggeri, Azzurra},
  journal={Human Behavior and Emerging Technologies},
  volume={3},
  number={4},
  pages={595--605},
  year={2021},
  publisher={Wiley Online Library}
}

@inproceedings{xu2020you,
  title={What are you talking to?: Understanding children's perceptions of conversational agents},
  author={Xu, Ying and Warschauer, Mark},
  booktitle={Proceedings of the 2020 CHI conference on human factors in computing systems},
  pages={1--13},
  year={2020},  
  publisher = {Association for Computing Machinery},
  address   = {New York, NY, USA},
  doi       = {10.1145/3313831.3376416}
}

@inproceedings{lovato2019hey,
  title={Hey Google, do unicorns exist? Conversational agents as a path to answers to children's questions},
  author={Lovato, Silvia B and Piper, Anne Marie and Wartella, Ellen A},
  booktitle={Proceedings of the 18th ACM international conference on interaction design and children},
  pages={301--313},
  year={2019},
  publisher = {Association for Computing Machinery},
  address   = {New York, NY, USA},
  doi       = {10.1145/3311927.3323150},
}

@inproceedings{druga2017hey,
  title={"Hey Google is it OK if I eat you?": Initial explorations in child-agent interaction},
  author={Druga, Stefania and Williams, Randi and Breazeal, Cynthia and Resnick, Mitchel},
  booktitle={Proceedings of the 2017 Conference on Interaction Design and Children},
  pages={595--600},
  year={2017},
  publisher = {Association for Computing Machinery},
  address   = {New York, NY, USA},
  doi       = {10.1145/3078072.3084330}
}

@inproceedings{belghith2024testing,
  title={Testing, socializing, exploring: Characterizing middle schoolers’ approaches to and conceptions of ChatGPT},
  author={Belghith, Yasmine and Mahdavi Goloujeh, Atefeh and Magerko, Brian and Long, Duri and Mcklin, Tom and Roberts, Jessica},
  booktitle={Proceedings of the 2024 CHI Conference on Human Factors in Computing Systems},
  pages={1--17},
  year={2024},
  publisher = {Association for Computing Machinery},
  address   = {New York, NY, USA},
  articleno = {276},
  numpages  = {17},
  doi       = {10.1145/3613904.3642332},
}

@inproceedings{dangol2025ai,
  title={“AI just keeps guessing”: Using ARC puzzles to help children identify reasoning errors in generative AI},
  author={Dangol, Aayushi and Zhao, Runhua and Wolfe, Robert and Ramanan, Trushaa and Kientz, Julie A and Yip, Jason},
  booktitle={Proceedings of the 24th interaction design and children},
  pages={444--464},
  year={2025},
  publisher = {Association for Computing Machinery},
  address   = {New York, NY, USA},
  doi       = {10.1145/3713043.3728836},
}

@misc{kosoy2024children,
  author        = {Kosoy, Eliza and Jeong, Soojin and Sinha, Anoop and Gopnik, Alison and Kraljic, Tanya},
  title         = {Children's Mental Models of Generative Visual and Text Based {AI} Models},
  year          = {2024},
  eprint        = {2405.13081},
  archivePrefix = {arXiv},
  primaryClass  = {cs.HC}
}

@inproceedings{oh2025hey,
  title={" Hey Curio, Can You Tell Me More?": Children's Information-Seeking and Trust in AI},
  author={Oh, Sunhyo and Zhang, Chao and Girouard-Hallam, Lauren and Zhou, Zihao and March, Halie and Jayaramu, Srushti and Xu, Ying},
  booktitle={Proceedings of the 24th interaction design and children},
  pages={545--555},
  year={2025},
  publisher = {Association for Computing Machinery},
  address   = {New York, NY, USA},
  doi       = {10.1145/3713043.3727059}
}

@article{vahedian2025ask,
  title={Ask me anything: exploring children’s attitudes toward an age-tailored AI-powered chatbot},
  author={Vahedian Movahed, Saniya and Martin, Fred G},
  journal={International Journal of Artificial Intelligence in Education},
  volume={35},
  number={6},
  pages={3979--4001},
  year={2025},
  publisher={Springer}
}

@article{xu2025makes,
  title={What makes children perceive or not perceive minds in generative AI?},
  author={Xu, Ying and Thomas, Trisha and Yu, Chi-Lin and Pan, Echo Zexuan},
  journal={Computers in Human Behavior: Artificial Humans},
  volume={4},
  pages={100135},
  year={2025},
  publisher={Elsevier}
}

@article{melson2009robotic,
  title={Robotic pets in human lives: Implications for the human--animal bond and for human relationships with personified technologies},
  author={Melson, Gail F and Kahn, Jr, Peter H and Beck, Alan and Friedman, Batya},
  journal={Journal of Social Issues},
  volume={65},
  number={3},
  pages={545--567},
  year={2009},
  publisher={Wiley Online Library}
}

@inproceedings{kahn2004robotic,
  title={Robotic pets in the lives of preschool children},
  author={Kahn Jr, Peter H and Friedman, Batya and Perez-Granados, Deanne R and Freier, Nathan G},
  booktitle={CHI'04 extended abstracts on Human factors in computing systems},
  pages={1449--1452},
  year={2004},
  publisher = {Association for Computing Machinery},
  address   = {New York, NY, USA},
  doi       = {10.1145/985921.986087},
}

@inproceedings{mcreynolds2017toys,
  title={Toys that listen: A study of parents, children, and internet-connected toys},
  author={McReynolds, Emily and Hubbard, Sarah and Lau, Timothy and Saraf, Aditya and Cakmak, Maya and Roesner, Franziska},
  booktitle={Proceedings of the 2017 CHI conference on human factors in computing systems},
  pages={5197--5207},
  year={2017},
  publisher = {Association for Computing Machinery},
  address   = {New York, NY, USA},
  doi       = {10.1145/3025453.3025735}
}

@article{steeves2020dialogic,
  title={A dialogic analysis of Hello Barbie’s conversations with children},
  author={Steeves, Valerie},
  journal={Big Data \& Society},
  volume={7},
  number={1},
  pages={2053951720919151},
  year={2020},
  publisher={SAGE Publications Sage UK: London, England}
}

@inproceedings{dangol2026toys,
  title     = {Toys that Listen, Talk, and Play: Understanding Children's Sensemaking and Interactions with {AI} Toys},
  author    = {Dangol, Aayushi and Gupta, Meghna and Yoo, Daeun and Wolfe, Robert and Yip, Jason and Roesner, Franziska and Kientz, Julie A.},
  booktitle = {Proceedings of the 25th Annual ACM Interaction Design and Children Conference},
  pages     = {818--834},
  year      = {2026},
  publisher = {Association for Computing Machinery},
  address   = {New York, NY, USA},
  doi       = {10.1145/3773077.3806142}
}

@techreport{goodacre2026ai,
  title={AI in the Early Years: Examining the Implications of GenAI Toys for Young Children},
  author={Goodacre, Emily and Gibson, Jenny},
  institution={University of Cambridge},
  year={2026},
  doi={10.17863/CAM.126270}
}

@article{mertala2020connectivity,
  title={How connectivity affects otherwise traditional toys? A functional analysis of hello Barbie},
  author={Mertala, Pekka},
  journal={International Journal of Child-Computer Interaction},
  volume={25},
  pages={100186},
  year={2020},
  publisher={Elsevier}
}

@inproceedings{hubbard2021voice,
  title={Voice design to support young children’s agency in child-agent interaction},
  author={Hubbard, Layne and Ding, Shanli and Le, Vananh and Kim, Pilyoung and Yeh, Tom},
  booktitle={Proceedings of the 3rd Conference on Conversational User Interfaces},
  pages={1--10},
  year={2021},
  publisher = {Association for Computing Machinery},
  address   = {New York, NY, USA},
  articleno = {9},
  numpages  = {10},
  doi       = {10.1145/3469595.3469604},
}

@inproceedings{druga2021perceptions,
  author    = {Druga, Stefania and Ko, Amy J.},
  title     = {How Do Children's Perceptions of Machine Intelligence Change When Training and Coding Smart Programs?},
  booktitle = {Proceedings of the 20th Annual ACM Interaction Design and Children Conference},
  pages     = {49--61},
  year      = {2021},
  publisher = {Association for Computing Machinery},
  doi       = {10.1145/3459990.3460712},
  address   = {New York, NY, USA}
}

@article{lin2020zhorai,
  author  = {Lin, Phoebe and Van Brummelen, Jessica and Lukin, Galit and Williams, Randi and Breazeal, Cynthia},
  title   = {{Zhorai}: Designing a Conversational Agent for Children to Explore Machine Learning Concepts},
  journal = {Proceedings of the AAAI Conference on Artificial Intelligence},
  volume  = {34},
  number  = {09},
  pages   = {13381--13388},
  year    = {2020},
  doi     = {10.1609/aaai.v34i09.7061}
}

@article{tian2023amby,
  author  = {Tian, Xiaoyi and Kumar, Amit and Solomon, Carly E. and Calder, Kaceja D. and Katuka, Gloria Ashiya and Song, Yukyeong and Celepkolu, Mehmet and Pezzullo, Lydia and Barrett, Joanne and Boyer, Kristy Elizabeth and Israel, Maya},
  title   = {{AMBY}: A Development Environment for Youth to Create Conversational Agents},
  journal = {International Journal of Child-Computer Interaction},
  volume  = {38},
  pages   = {100618},
  year    = {2023},
  doi     = {10.1016/j.ijcci.2023.100618}
}

@inproceedings{patton2025aptly,
  author    = {Patton, Evan W. and Kim, David Y. J. and Granquist, Ashley M. and Liu, Robin and Scott, Arianna and Zamanova, Jennet and Abelson, Harold},
  title     = {Aptly: Making Mobile Apps from Natural Language},
  booktitle = {Extended Abstracts of the CHI Conference on Human Factors in Computing Systems},
  articleno = {73},
  pages     = {1--6},
  year      = {2025},
  publisher = {Association for Computing Machinery},
  doi       = {10.1145/3706599.3720081},
  address   = {New York, NY, USA},
  numpages  = {6}
}

@inproceedings{tian2026hallucination,
  author    = {Tian, Xiaoyi and Ozturk, Deniz and Edula, Sreekar and Adil, Jibran and Jin, Qiao and Shi, Yang and Barnes, Tiffany},
  title     = {When {AI} Gets It Wrong: Scaffolding {AI} Hallucination Detection for Children Through Chatbot Creation},
  booktitle = {Proceedings of the 2026 CHI Conference on Human Factors in Computing Systems},
  pages     = {1--20},
  year      = {2026},
  publisher = {Association for Computing Machinery},
  doi       = {10.1145/3772318.3791480},  
  address   = {New York, NY, USA}
}

@inproceedings{bai2023participatory,
  author    = {Bai, Zhen and Judd, Frances and Polinsky, Naomi and Yadollahi, Elmira},
  title     = {Participatory Design of {AI} with Children: Reflections on {IDC} Design Challenge},
  booktitle = {CHI 2023 Workshop on Child-centred AI Design: Definition, Operation and Considerations},
  year      = {2023},
  publisher = {Association for Computing Machinery},
  address   = {New York, NY, USA},
  location  = {Hamburg, Germany},
  numpages  = {4},
  url       = {https://arxiv.org/abs/2304.09091}
}

@article{wilson2025childcentred,
  author  = {Wilson, Cara and Atabey, Ay{\c{c}}a and Revans, Joe},
  title   = {Towards Child-Centred {AI} in Children's Learning Futures: Participatory Design Futuring with {SmartSchool} and the Co-Design Stories Toolkit},
  journal = {International Journal of Human-Computer Studies},
  volume  = {199},
  pages   = {103431},
  year    = {2025},
  doi     = {10.1016/j.ijhcs.2024.103431}
}

@inproceedings{alvesoliveira2021children,
  author    = {Alves-Oliveira, Patr{\'i}cia and Arriaga, Patr{\'i}cia and Paiva, Ana and Hoffman, Guy},
  title     = {Children as Robot Designers},
  booktitle = {Proceedings of the 2021 ACM/IEEE International Conference on Human-Robot Interaction},
  address   = {New York, NY, USA},
  pages     = {399--408},
  year      = {2021},
  publisher = {Association for Computing Machinery},
  doi       = {10.1145/3434073.3444650}
}

@article{neerincx2023codesigning,
  author  = {Neerincx, Anouk and Veldhuis, Daan and Masthoff, Judith M. F. and de Graaf, Maartje M. A.},
  title   = {Co-designing a Social Robot for Child Health Care},
  journal = {International Journal of Child-Computer Interaction},
  volume  = {38},
  pages   = {100615},
  year    = {2023},
  doi     = {10.1016/j.ijcci.2023.100615}
}

@article{levinson2025think,
  author  = {Levinson, Leigh and Charisi, Vicky and Zotos, Chris and Gomez, Randy and {\v{S}}abanovi{\'c}, Selma},
  title   = {Let Us Make Robots ``Think in Child!'': How Children Conceptualize Fairness, Inclusion, and Privacy with Social Robots},
  journal = {International Journal of Child-Computer Interaction},
  volume  = {43},
  pages   = {100706},
  year    = {2025},
  doi     = {10.1016/j.ijcci.2024.100706}
}

@inproceedings{zhou2024lollipop,
  author    = {Zhou, Hanqing and Nikolova, Anastasia and An, Pengcheng},
  title     = {{``My Lollipop Dropped...''---Probing Design Opportunities for SEL Agents through Children's Peer Co-Creation of Social-Emotional Stories}},
  booktitle = {Extended Abstracts of the CHI Conference on Human Factors in Computing Systems},
  articleno = {12},
  pages     = {1--8},
  year      = {2024},
  publisher = {Association for Computing Machinery},
  doi       = {10.1145/3613905.3651867},
  address   = {New York, NY, USA},
  numpages  = {8}
}

@incollection{vanmechelen2019internet,
  author    = {Van Mechelen, Maarten and Zaman, Bieke and Bleumers, Lizzy and Mari{\"e}n, Ilse},
  title     = {Designing the Internet of Toys for and with Children: A Participatory Design Case Study},
  booktitle = {The Internet of Toys: Practices, Affordances and the Political Economy of Children's Smart Play},
  editor    = {Mascheroni, Giovanna and Holloway, Donell},
  series    = {Studies in Childhood and Youth},
  pages     = {181--203},
  publisher = {Palgrave Macmillan},
  address   = {Cham},
  year      = {2019},
  doi       = {10.1007/978-3-030-10898-4_9}
}

@incollection{yamadarice2019including,
  author    = {Yamada-Rice, Dylan},
  title     = {Including Children in the Design of the Internet of Toys},
  booktitle = {The Internet of Toys: Practices, Affordances and the Political Economy of Children's Smart Play},
  editor    = {Mascheroni, Giovanna and Holloway, Donell},
  series    = {Studies in Childhood and Youth},
  pages     = {205--221},
  publisher = {Palgrave Macmillan},
  address   = {Cham},
  year      = {2019},
  doi       = {10.1007/978-3-030-10898-4_10}
}

@inproceedings{cai2025childai,
  author    = {Cai, Zhenyao and Han, Ariel and Zhou, Xiaofei and Durall Gazulla, Eva and Peppler, Kylie},
  title     = {Child-{AI} Co-Creation: A Review of the Current Research Landscape and a Proposal for Six Design Considerations},
  booktitle = {Proceedings of the 24th Interaction Design and Children Conference},
  pages     = {916--922},
  year      = {2025},
  publisher = {Association for Computing Machinery},
  doi       = {10.1145/3713043.3731506},
  address   = {New York, NY, USA},
}

@inproceedings{zhang2024mathemyths,
  author    = {Zhang, Chao and Liu, Xuechen and Ziska, Katherine and Jeon, Soobin and Yu, Chi-Lin and Xu, Ying},
  title     = {Mathemyths: Leveraging Large Language Models to Teach Mathematical Language through Child-{AI} Co-Creative Storytelling},
  booktitle = {Proceedings of the 2024 CHI Conference on Human Factors in Computing Systems},
  articleno = {274},
  pages     = {1--23},
  year      = {2024},
  publisher = {Association for Computing Machinery},
  doi       = {10.1145/3613904.3642647},  
  address   = {New York, NY, USA},
  numpages  = {23}
}

@inproceedings{chen2024chatscratch,
  author    = {Chen, Liuqing and Xiao, Shuhong and Chen, Yunnong and Wu, Ruoyu and Song, Yaxuan and Sun, Lingyun},
  title     = {ChatScratch: An {AI}-Augmented System Toward Autonomous Visual Programming Learning for Children Aged 6--12},
  booktitle = {Proceedings of the 2024 CHI Conference on Human Factors in Computing Systems},
  articleno = {649},
  pages     = {1--19},
  year      = {2024},
  publisher = {Association for Computing Machinery},
  doi       = {10.1145/3613904.3642229},
  address   = {New York, NY, USA},
  numpages  = {19}
}

@inproceedings{newman2024darth,
  author    = {Newman, Michele and Sun, Kaiwen and Dalla Gasperina, Ilena B. and Shin, Grace Y. and Pedraja, Matthew Kyle and Kanchi, Ritesh and Song, Maia B. and Li, Rannie and Lee, Jin Ha and Yip, Jason},
  title     = {{``I Want It to Talk Like Darth Vader'': Helping Children Construct Creative Self-Efficacy with Generative AI}},
  booktitle = {Proceedings of the 2024 CHI Conference on Human Factors in Computing Systems},
  articleno = {117},
  pages     = {1--18},
  year      = {2024},
  publisher = {Association for Computing Machinery},
  doi       = {10.1145/3613904.3642492},
  address   = {New York, NY, USA},
  numpages  = {18}
}

@article{xiao2025intelligent,
  title={Intelligent toys, complex questions: A literature review of artificial intelligence in children's toys and devices},
  author={Xiao, Wei and Gon{\c{c}}alves, Alexandre},
  journal={Big Data \& Society},
  volume={12},
  number={4},
  pages={20539517251389860},
  year={2025},
  publisher={SAGE Publications Sage UK: London, England}
}

@article{palaiologou2021make,
  title={Make-believe play with the Internet of Toys: A case for multimodal playscapes},
  author={Palaiologou, Ioanna and Kewalramani, Sarika and Dardanou, Maria},
  journal={British Journal of Educational Technology},
  volume={52},
  number={6},
  pages={2100--2117},
  year={2021},
  publisher={Wiley Online Library}
}

@article{akdeniz2021maya,
  title={Maya: An artificial intelligence based smart toy for pre-school children},
  author={Akdeniz, Mevl{\"u}de and {\"O}zdin{\c{c}}, Fatih},
  journal={International Journal of Child-Computer Interaction},
  volume={29},
  pages={100347},
  year={2021},
  publisher={Elsevier}
}

@inproceedings{hutchinson2003technology,
  title={Technology probes: inspiring design for and with families},
  author={Hutchinson, Hilary and Mackay, Wendy and Westerlund, Bo and Bederson, Benjamin B and Druin, Allison and Plaisant, Catherine and Beaudouin-Lafon, Michel and Conversy, St{\'e}phane and Evans, Helen and Hansen, Heiko and others},
  booktitle={Proceedings of the SIGCHI conference on Human factors in computing systems},
  pages={17--24},
  year={2003},
  publisher = {Association for Computing Machinery},
  address   = {New York, NY, USA},
  doi       = {10.1145/642611.642616},
}

@inproceedings{long2020ai,
  title={What is AI literacy? Competencies and design considerations},
  author={Long, Duri and Magerko, Brian},
  booktitle={Proceedings of the 2020 CHI conference on human factors in computing systems},
  pages={1--16},
  year={2020},
  publisher = {Association for Computing Machinery},
  address   = {New York, NY, USA},
  doi       = {10.1145/3313831.3376727}
}

@article{sundar2010personalization,
  author  = {Sundar, S. Shyam and Marathe, Sampada S.},
  title   = {Personalization versus Customization: The Importance of Agency, Privacy, and Power Usage},
  journal = {Human Communication Research},
  year    = {2010},
  volume  = {36},
  number  = {3},
  pages   = {298--322},
  doi     = {10.1111/j.1468-2958.2010.01377.x}
}

@incollection{sundar2008self,
  author    = {Sundar, S. Shyam},
  title     = {Self as Source: Agency and Customization in Interactive Media},
  booktitle = {Mediated Interpersonal Communication},
  editor    = {Konijn, Elly A. and Utz, Sonja and Tanis, Martin and Barnes, Susan B.},
  year      = {2008},
  pages     = {58--74},
  publisher = {Routledge},
  address   = {New York},
  isbn      = {9780805863048}
}

@article{kang2016self,
  author  = {Kang, Hyunjin and Sundar, S. Shyam},
  title   = {When Self Is the Source: Effects of Media Customization on Message Processing},
  journal = {Media Psychology},
  year    = {2016},
  volume  = {19},
  number  = {4},
  pages   = {561--588},
  doi     = {10.1080/15213269.2015.1121829}
}

@inproceedings{jones2025agentmemory,
  author    = {Jones, Brennan and Stemmler, Kelsey and Su, Emily and Kim, Young-Ho and Kuzminykh, Anastasia},
  title     = {Users' Expectations and Practices with Agent Memory},
  booktitle = {Extended Abstracts of the CHI Conference on Human Factors in Computing Systems},
  year      = {2025},
  publisher = {Association for Computing Machinery},
  address   = {New York, NY, USA},
  pages     = {1--8},
  doi       = {10.1145/3706599.3720158},
  isbn      = {9798400713958}
}

@inproceedings{jiang2026recallbot,
  author    = {Jiang, Zhaojun and Zheng, Chunyuan and Chen, Hongyi and Chen, Liuqing},
  title     = {{RECALLbot}: Designing Agentic Memory and Reciprocal Disclosure for Human--Chatbot Relationships},
  booktitle = {Proceedings of the 2026 CHI Conference on Human Factors in Computing Systems},
  year      = {2026},
  publisher = {Association for Computing Machinery},
  address   = {New York, NY, USA},
  articleno = {795},
  pages     = {1--20},
  doi       = {10.1145/3772318.3790714},
  numpages  = {20}
}

@article{do2023inform,
  author  = {Do, Hyo Jin and Kong, Ha-Kyung and Tetali, Pooja and Karahalios, Karrie and Bailey, Brian P.},
  title   = {Inform, Explain, or Control: Techniques to Adjust End-User Performance Expectations for a Conversational Agent Facilitating Group Chat Discussions},
  journal = {Proceedings of the ACM on Human-Computer Interaction},
  year    = {2023},
  volume  = {7},
  number  = {CSCW2},
  articleno = {343},
  pages   = {1--26},
  doi     = {10.1145/3610192},
  numpages= {26}
}

@article{williams2019popbots,
  author  = {Williams, Randi and Park, Hae Won and Oh, Lauren and Breazeal, Cynthia},
  title   = {{PopBots}: Designing an Artificial Intelligence Curriculum for Early Childhood Education},
  journal = {Proceedings of the AAAI Conference on Artificial Intelligence},
  year    = {2019},
  volume  = {33},
  number  = {01},
  pages   = {9729--9736},
  doi     = {10.1609/aaai.v33i01.33019729}
}

@inproceedings{dangol2025mentalmodels,
  author    = {Dangol, Aayushi and Wolfe, Robert and Zhao, Runhua and Kim, JaeWon and Ramanan, Trushaa and Davis, Katie and Kientz, Julie A.},
  title     = {Children's Mental Models of {AI} Reasoning: Implications for {AI} Literacy Education},
  booktitle = {Proceedings of the 24th Interaction Design and Children Conference},
  year      = {2025},
  pages     = {106--123},
  publisher = {Association for Computing Machinery},
  address   = {New York, NY, USA},
  doi       = {10.1145/3713043.3728856}
}

@inproceedings{malnatsky2026making,
  author    = {Malnatsky, Elena and Wang, Shenghui and Hindriks, Koen and Ligthart, Mike E. U.},
  title     = {Making Memories Together: Toward Transparent and Co-constructed Robot Memory in Child-Robot Interaction},
  booktitle = {Companion Proceedings of the 21st ACM/IEEE International Conference on Human-Robot Interaction},
  pages     = {1125--1129},
  year      = {2026},
  publisher = {Association for Computing Machinery},
  address   = {New York, NY, USA},
  doi       = {10.1145/3776734.3794570}
}

@article{frazier2009preschoolers,
  author  = {Frazier, Brandy N. and Gelman, Susan A. and Wellman, Henry M.},
  title   = {Preschoolers' Search for Explanatory Information Within Adult--Child Conversation},
  journal = {Child Development},
  year    = {2009},
  volume  = {80},
  number  = {6},
  pages   = {1592--1611},
  doi     = {10.1111/j.1467-8624.2009.01356.x}
}

@article{yuan2019reformulations,
  author  = {Yuan, Ye and Thompson, Stryker and Watson, Kathleen and Chase, Alice and Senthilkumar, Ashwin and Brush, A. J. Bernheim and Yarosh, Svetlana},
  title   = {Speech Interface Reformulations and Voice Assistant Personification Preferences of Children and Parents},
  journal = {International Journal of Child-Computer Interaction},
  year    = {2019},
  volume  = {21},
  pages   = {77--88},
  doi     = {10.1016/j.ijcci.2019.04.005}
}

@article{vartiainen2020learning,
  author  = {Vartiainen, Henriikka and Tedre, Matti and Valtonen, Teemu},
  title   = {Learning Machine Learning with Very Young Children: Who Is Teaching Whom?},
  journal = {International Journal of Child-Computer Interaction},
  year    = {2020},
  volume  = {25},
  pages   = {100182},
  doi     = {10.1016/j.ijcci.2020.100182}
}

@inproceedings{dwivedi2021exploring,
  author    = {Dwivedi, Utkarsh and Gandhi, Jaina and Parikh, Raj and Coenraad, Merijke and Bonsignore, Elizabeth and Kacorri, Hernisa},
  title     = {Exploring Machine Teaching with Children},
  booktitle = {2021 IEEE Symposium on Visual Languages and Human-Centric Computing (VL/HCC)},
  pages     = {1--11},
  year      = {2021},
  publisher = {IEEE},
  doi       = {10.1109/VL/HCC51201.2021.9576171},
  address   = {Piscataway, NJ, USA}
}

@article{norton2012ikea,
  title={The IKEA effect: When labor leads to love},
  author={Norton, Michael I and Mochon, Daniel and Ariely, Dan},
  journal={Journal of consumer psychology},
  volume={22},
  number={3},
  pages={453--460},
  year={2012},
  publisher={Wiley Online Library}
}

@phdthesis{druga2018growing,
  title={Growing up with AI: Cognimates: from coding to teaching machines},
  author={Druga, Stefania},
  year={2018},
  school={Massachusetts Institute of Technology}
}

@article{misirli2023computational,
  author  = {Misirli, Anastasia and Komis, Vassilis},
  title   = {Computational thinking in early childhood education: The impact of programming a tangible robot on developing debugging knowledge},
  journal = {Early Childhood Research Quarterly},
  volume  = {65},
  pages   = {139--158},
  year    = {2023},
  doi     = {10.1016/j.ecresq.2023.05.014}
}

\end{document}